\documentclass{aa}  

\usepackage{graphicx}
\usepackage{txfonts}

\usepackage[colorlinks=true,linkcolor=blue,citecolor=blue,urlcolor=blue]{hyperref}
\usepackage{multicol}
\begin{document}

   \title{Improving 1D stellar atmosphere models with insights from multi-dimensional simulations}

   \subtitle{II. 1D versus 3D hydrodynamically consistent model comparison for WR stars}

\author{G.\ Gonz\'alez-Tor\`a
          \inst{\ref{inst:ARI}}
          \and
          A.\ A.\ C.\ Sander\inst{\ref{inst:ARI}}
          \and 
          N.\ Moens\inst{\ref{inst:KUL}}
        \and
        J.\ O.\ Sundqvist\inst{\ref{inst:KUL}}
        \and
        D.\ Debnath\inst{\ref{inst:KUL}}
        \and 
        L.\ Delbroek\inst{\ref{inst:KUL}}
        \and 
        J.\ Josiek\inst{\ref{inst:ARI}}
        \and 
        R.\ R.\ Lefever\inst{\ref{inst:ARI}}
        \and 
        C.\ Van der Sijpt\inst{\ref{inst:KUL}}
        \and
        O.\ Verhamme\inst{\ref{inst:KUL}}
        \and 
        M.\ Bernini-Peron\inst{\ref{inst:ARI}}
        }

   \institute{{Zentrum für Astronomie der Universität Heidelberg, Astronomisches Rechen-Institut, 
   Mönchhofstr. 12-14, 69120 Heidelberg\label{inst:ARI}}\\
        \email{gemma.gonzalez-tora@uni-heidelberg.de}
   \and
        {Institute of Astronomy, KU Leuven, Celestijnenlaan 200D, 3001, Leuven, Belgium\label{inst:KUL}}\\
}  

   \date{Received \today; accepted -}

% \abstract{}{}{}{}{} 
% 5 {} token are mandatory
 
  \abstract
   {Classical Wolf-Rayet (cWR) stars are evolved massive stars that have lost most of their hydrogen envelope, presenting dense, extended atmospheres with strong stellar winds. Accurate descriptions of their line-driven winds and in particular the launching of the winds in optically thick layers have long remained enigmatic. Two different approaches have recently allowed to make significant progress, namely one-dimensional atmosphere models with stationary hydrodynamics and time-dependent, multi-dimensional, radiation-hydrodynamic models.
   }
   {The computational demands and required approximations limit the applications of multi-dimensional, time-dependent atmospheric models. 1D stationary atmosphere models therefore remain an important and necessary tool, but need to incorporate reasonable approximations of the physical insights gained in multi-dimensional and time-dependent simulations. 
   }
   {
   We compare averaged stratifications from recent multi-dimensional atmosphere models for cWR stars with 1D stellar atmosphere models computed with the hydrodynamically consistent branch of the PoWR model atmosphere code. We study models with winds launched by the hot iron bump and vary several of the 1D model parameters to characterize occurring differences and quantify their impact on the obtained 1D solutions.  
   }
   {The 1D hydrodynamically consistent atmosphere models match the averaged 3D density stratifications. 1D models with standard input obtain mass-loss rates $\lesssim0.2$ dex higher than their 3D equivalent, but small adjustments in accordance with the mass-loss and luminosity dispersion obtained from the time-dependent multi-dimensional simulations can reconcile this difference. 
   The 1D models are radially more extended, show higher terminal velocities and lower effective temperatures. 
   While the 1D models reproduce the same velocity trend as the 3D calculations, they launch their winds a bit further out and reaching higher velocities during the hot iron bump. The resulting differences in effective temperature and terminal velocity are also seen in the synthetic spectra computed from different 1D PoWR model approaches.   
   }
   {The overall stellar atmosphere structure profiles for 1D and 3D averaged hydrodynamically consistent models follow a similar trend, obtaining stellar parameters that are well matched when accounting for the dispersion of the time-dependent 3D simulations and the different methodologies. For stars closer to the Eddington Limit, decreasing the Doppler velocities in 1D models can help reconcile the mass-loss rates, effective temperatures, and velocity profiles in the outer wind. Obtaining matching temperature structures in optically thin regions remains an open challenge.
   }

   
   \maketitle
%
%-------------------------------------------------------------------

\section{Introduction}\label{sec:intro}

Massive stars ($M_\text{init}\gtrsim8\,M_{\odot}$) and their stellar winds play a crucial role in the chemical enrichment of galaxies \citep[e.g.,][]{Portinari+1998}, not only by ionizing the interstellar medium (ISM) but also by enriching the ISM with nuclear processed material. The strong, UV-dominated radiation field drives a powerful stellar wind by transferring momentum from photons via absorption and scattering in metal line transitions \citep[e.g.,][]{Castor+1975,Pauldrach+1986}. Before ending their lives, some massive stars can completely deplete their hydrogen, even in their outer layers, and become a ``stripped'' helium star. This stripping can happen via strong, intrinsic stellar winds \citep{Conti1975,Abbott1987}, binary interaction \citep{Paczynski1967,Shenar+2020,Dsilva+2020}, or eruptive events \citep{Smith2014}. Stars showing spectra dominated by strong, broad emission lines, are classified as Wolf-Rayet (WR) stars \citep[named after][]{WolfRayet1867}. If these stars are also He burning, usually coinciding with highly depleted or absent hydrogen envelopes, they are called ``classical''\footnote{Historically, the term ``classical WR'' was used for all Population I WR stars \citep[e.g.,][]{vanderHucht+1981}. This slowly changed after the discovery of hydrogen-burning very massive stars with WR-type spectral features \citep[e.g.,][]{deKoter+1997,SmithConti2008}} WR stars \citep[e.g.,][]{Moffat2015,Shenar2024}.

Developing accurate stellar atmospheres to reproduce the spectra of WR stars is vital to understand the underlying physical processes inherent in these massive stars. Due to their intense radiation, low densities, and supersonic velocity fields, usual assumptions such as local thermodynamical equilibrium (LTE) cannot be applied and the radiative transfer needs to account for specific line transitions as well as the expanding atmosphere, which is numerically challenging \citep[see e.g.,][]{Hillier2003,HubenyLanz2003,HamannGraefener2003}. Consequently, non-LTE atmosphere modeling for the spectral analysis of O, B, and WR stars is essentially performed only via 1D stationary models \citep[see, e.g.,][]{Puls2008,Sander2017}, typically solving the radiative transfer equations in the co-moving frame (CMF), using state-of-the-art expanding atmosphere codes such as CMFGEN \citep{HillierMiller1998}, PoWR \citep{Graefener+2002}, and FASTWIND \citep{Puls+2005}. A recent overview and analysis method comparison for O stars is given in \citet{Sander+2024}.

Despite the existence of modern model atmosphere codes, understanding classical WR stars has turned out to be particularly challenging. One of the reasons is the so-called ``WR radius problem'' referring to significant inconsistencies between the empirically inferred stellar radii and the predictions from stellar structure models \citep[see, e.g.,][]{Hamann+2006,Graefener+2012}. Traditionally, the wind regime in atmosphere models used for quantitative spectroscopy is described by adopting a $\beta$-law velocity profile. For stars with dense stellar winds, the full emerging spectrum, including the continuum, is produced in the outer wind, leading to a significant parameter degeneracy with respect to the inferred radii \citep[see, e.g., the discussion in][]{HamannGraefener2004}, which becomes even larger when allowing for different $\beta$ values \citep{Lefever+2023}. As demonstrated by \citet{GraefenerHamann2005} and later also \citet{Sander+2020} and \citet{Poniatowski+2021}, the radius problem can in principle be resolved if the velocity and density stratification is instead calculated consistent from the hydrodynamic equation of motion. Full hydrodynamic models are therefore particularly crucial to understand the stellar atmospheres and spectra of WR stars. The concept of 1D hydrodynamically-consistent CMF models goes back to \citet{Pauldrach+1986}, and was first applied to WR stars in \citet{GraefenerHamann2005,GraefenerHamann2008}. Later, a dedicated hydrodynamically-consistent PoWR branch (PoWR$^{\textsc{hd}}$) has been developed \citep{Sander+2017,Sander+2023}, which we also employ in this work. 

 Pioneering work on time-dependent, multi-dimensional, radiation-hydrodynamical (RHD) simulations with a hybrid opacity approach \citep{Moens+2022fld,Poniatowski+2022} for classical WR stars was carried out by \citet{Moens+2022wr}, later extended to the O star regime by \citet{Debnath+2024,ud-Doula+2025}. \citet{Moens+2022wr} presented three cWR atmosphere models with luminosities of $\log\,L/L_{\odot}=5.47, 5.64$, and $5.74$ (termed $\Gamma$2, $\Gamma$3, and $\Gamma$4, respectively). These RHD models launch an optically-thick supersonic wind from deep, sub-surface regions and are initiated due to the (hot) iron opacity peak. A fourth, lower luminosity model ($\log\,L/L_{\odot}=5.33$, denoted as $\Gamma$1) from \citet{Moens+2022wr} shows instead a more standard, line-driven wind launched from higher, optically thin atmospheric layers.

 In the first paper of this series \citep{GonzalezTora+25}, we compared the multi-dimensional RHD models by \citet{Debnath+2024} with ``standard'' 1D PoWR models constraining the (quasi-)hydrostatic regime and a $\beta$-law to describe the stellar wind region. \citet{GonzalezTora+25} demonstrated that the inclusion of a turbulent pressure in the solution of the hydrostatic equation accurately reproduces the averaged density from the 2D models by \citet{Debnath+2024}. The velocity and partially also the temperature stratification were reproduced reasonably well, even when just assuming a constant turbulent velocity throughout the atmosphere. Yet, these models were limited to O stars, where 1D hydrodynamic results can be reasonably well approximated by a $\beta$-law \citep[e.g.,][]{GraefenerHamann2008,Hamann+2008}\footnote{This does not apply for the crucial (average) wind launching region around the sonic point \citep{Sander+2017,GonzalezTora+25}}.
 In this work, we shift the focus to WR stars and compare the 3D model results for the three highest luminosity cWR stellar models from \citet{Moens+2022wr} with new, 1D hydrodynamically-consistent PoWR$^{\textsc{hd}}$ model calculations.

This paper is organized as follows: In Section~\ref{sec:methods} we present the main characteristics of the the hydrodynamically-consistent 1D models with PoWR$^{\textsc{hd}}$ and highlight the main differences compared to the stratification derived in the multi-dimensional hydrodynamical framework. Section~\ref{sec:results} shows the profile comparisons for both modelling approaches and discusses the implications of our results, and conclude our work in Section~\ref{sec:conc}.

%--------------------------------------------------------------------

 \section{Methods}\label{sec:methods}

\subsection{The PoWR$^{\,\textsc{hd}}$ 1D model}
The Potsdam Wolf-Rayet stellar atmosphere code \citep[PoWR,][]{Graefener+2002,HamannGraefener2003,Sander+2015} solves the CMF radiative transfer equations for a 1D spherical, stationary outflow. The non-LTE population numbers are calculated assuming statistical equilibrium \citep[e.g.,][]{Hamann1986} and the temperature stratification is obtained from a generalized Uns\"old-Lucy method \citep{HamannGraefener2003} or the thermal balance of electrons \citep{Kubat+1999,Sander2015}.

The inner boundary in PoWR models is set at a fixed Rosseland continuum optical depth $\tau_{\mathrm{Ross,cont}}$. 
Both the radius $R_\ast$ and temperature $T_\ast$ are defined at the same $\tau_{\mathrm{Ross,cont}}$, thereby differing from the usual effective temperature $T_{\mathrm{eff}}$ definition at $\tau_{\mathrm{Ross}}=2/3$. To avoid confusion and enable easier comparison with the 3D RHD models, we will label $T_{\mathrm{eff}} \equiv T_{2/3} := T(\tau_{\mathrm{Ross}}=2/3)$ in this work. The values of $T_\ast$ are less insightful here as they would be in traditional WR spectral analysis efforts \citep[e.g.,][]{Hamann+2006,Sander+2014} as we have adjusted the value of $\tau_{\mathrm{Ross,cont}}$ for the inner boundary in this work differently for each type of 1D model. This effort requires usually a small iteration of models, but is made to align our total Rosseland optical depth $\tau_{\mathrm{Ross}}$ at the inner boundary with the value from the corresponding 3D RHD simulations. 

There are two branches of PoWR which differ in their hydrodynamic treatment: In the standard branch, only the subsonic regime is solved consistently by integrating the hydrostatic equation. This equation can include an optional turbulent velocity term ($\varv_{\mathrm{turb}}$) which adds an extra turbulent pressure, $P_\mathrm{turb}(r) = \rho(r) \varv_{\mathrm{turb}}^{2}(r)$ \citep[see Eq.\,4 in][]{GonzalezTora+25}.
To account for the wind dynamics in the supersonic regime, the code adopts a $\beta$-law velocity profile
\begin{equation}
  \label{eq:vbeta}
    \varv(r)=p_{1}\left( 1-\frac{1}{r+p_{2}} \right)^\beta
\end{equation}
with the parameters $p_{1}$ and $p_{2}$ fixed by the boundary conditions $\varv(r_{\mathrm{max}})=\varv_{\infty}$ and $\varv(r_{\mathrm{con}})=\varv_{\mathrm{con}}$. Here, $r_{\mathrm{con}}$ is the connection point between the hydrostatic and the wind regime, which can either be set by demanding a smooth velocity gradient connection or by specifying a desired ratio $f_\text{sonic}$ between gas velocity and sound speed with values typically chosen between $0.5$ and unity. The density stratification $\rho(r)$ is then obtained from a given mass-loss rate $\dot{M}$ and the equation of continuity: $\dot{M} = 4 \pi r^2 \varv(r) \rho(r)$.
The standard, $\beta$-law approach has been used and discussed in \citet{GonzalezTora+25} in the context of O star modelling. For comparison, we also employ it in Sect.~\ref{sec:results}, where we show the differences with respect to PoWR$^{\textsc{hd}}$ solutions. 

One problem with the standard framework is that its treatment is not guaranteed to be self-consistent with respect to the wind hydrodynamics, neither locally nor globally. As $\dot{M}$ is a free input parameter to the models, the modeled wind might be stronger or weaker than what could actually be driven, meaning that such models have no predictive power with respect to the wind parameters. In addition, as disussed in Sect.\,\ref{sec:intro}, the restriction to $\beta$-type velocity laws is usually a sufficient representation for the averaged wind stratification in O stars \citep{GonzalezTora+25}, but insufficient for many cWR stars \citep[][]{GraefenerHamann2005,Sander+2020}.

The solution is to use a hydrodynamically consistent treatment, using the radiative acceleration from detailed radiative transfer and solving the full hydrodynamic equations \citep{Sander+2017,Sander+2023}.  The hydrodynamically consistent modelling approach is numerically very expensive, limiting the models to 1D, so multi-dimensional effects can only be parameterized.

The PoWR$^{\textsc{hd}}$ branch creates hydrodynamically-consistent atmosphere models by using the radiative acceleration $a_{\mathrm{rad}}$ from CMF radiative transfer and solving the stationary hydrodynamic equation in 1D 
\begin{equation}\label{eq:hydro}
    \varv\frac{\mathrm{d}\varv}{\mathrm{d}r}+\frac{GM_{\star}}{r^{2}}=a_{\mathrm{rad}}(r)+a_{\mathrm{press}}(r)
\end{equation}
\citep[e.g.,][]{Sander+2017} where $a_{\mathrm{press}}(r)$ is the acceleration due to gas (and optionally turbulent) pressure $P$, defined as 
\begin{equation}\label{eq:apress}
    a_{\mathrm{press}}(r):=-\frac{1}{\rho}\frac{\mathrm{d}P}{\mathrm{d}r}\text{.}
\end{equation}
If we rewrite the $a_{\mathrm{press}}-$term and replace all explicit $\rho$-dependencies \citep[see][for a detailed calculation]{Sander+2015}, Eq.\,\eqref{eq:hydro} can be rewritten in a compact way by defining the dimensionless quantities $\tilde{\mathcal{F}}$ and $\tilde{\mathcal{G}}$ \citep[see][]{Sander+2017}
\begin{equation}\label{eq:finalhydro}
    \frac{\mathrm{d}\varv}{\mathrm{d}r}=-\frac{g}{\varv}\frac{\tilde{\mathcal{F}}(r, M)}{\tilde{\mathcal{G}}(r, \varv)}\text{.}
\end{equation}
In Eq.\,\eqref{eq:finalhydro}, the radius dependencies of $\tilde{\mathcal{F}}$ and $\tilde{\mathcal{G}}$ are not just explicit, but also reflect implicit dependencies on $a_\text{rad}(r)$ (in the case of $\tilde{\mathcal{F}}$) and $a_\text{s}(r)$, with the latter denoting the square root of the sum of speed of sound and turbulent velocity squared \citep[see, e.g.,][]{GonzalezTora+25}. 
With the radiative acceleration described as a function of radius $a_{\mathrm{rad}}(r)$, Eq.\,\eqref{eq:finalhydro} has a critical point at $\tilde{\mathcal{G}} = 0$, which is also the sonic point, i.e., the point where the gas velocity $\varv$ is equal to $a_\text{s}$. In case of a non-zero turbulent velocity, the critical point shifts and the velocity has to be equal to the root of the sound speed squared plus the turbulent velocity squared. 
A starting model, e.g., a converged $\beta$-law model with a (quasi-)hydrostatic regime, is used as a first approximation of $a_{\mathrm{rad}}(r)$ to calculate the terms $\tilde{\mathcal{F}}(r)$ and $\tilde{\mathcal{G}}(r)$. Eq.\,\eqref{eq:finalhydro} is then integrated inwards and outwards from the critical point to obtain a consistent velocity field $\varv(r)$ and update $\dot{M}$ via the additional constraint to conserve the total $\tau_{\mathrm{Ross,cont}}$. %\jon{question (for us internally): this is down to a *fixed* lower boundary radius, I suppose?}
This additional correction step is added to the overall iteration scheme of the atmosphere calculations with further updates of $\dot{M}$ and $\varv(r)$ triggered until an overall convergence for the population numbers, the flux consistency, $\dot{M}$, $\varv_\infty$, and the conservation of the total $\tau_{\mathrm{Ross,cont}}$ is reached.   

Due to this additional iteration scheme, PoWR$^{\textsc{hd}}$ models are able to predict the $\dot{M}$ and $\varv(r)$ instead of having them as a free input \citep[see, e.g.,][]{SanderVink2020}. Yet, such a treatment is numerically very expensive, limiting it to 1D for the foreseeable future, and cannot handle non-monotonic velocity fields in the CMF radiative transfer. Note that we can also infer the luminosity $L$ or mass $M$ from a PoWR$^{\textsc{hd}}$ model if we instead fix $\dot{M}$. This approach is applied in Sect.~\ref{sec:powrs}.

Similarly to \citet{GonzalezTora+25}, we neglect potential effects from energy transport by enthalpy (`convection') in our 1D models. For the WR star models in this work, the convective energy transport reaches less than $10\,\%$ of the total luminosity at the innermost layers and decreases outwards even further. Therefore, our assumption of a purely radiative energy transport for the 1D PoWR models is well justified in this regime.

Both $\beta$-law PoWR and PoWR$^{\textsc{hd}}$ models account for wind inhomogeneities using the \textit{microclumping} approach assuming small scale, optically thin clumps surrounded by a void interclump medium. The ``clumping factor'' in PoWR is specified as a density contrast $D(r)$ describing the density enhancement of the clumps compared to a smooth wind with the same mass-loss rate \citep{HamannKoesterke1998}. In this work, we explored the effect of different density contrasts in Sect.\,\ref{sec:powrs} for 1D representations of the $\Gamma$3 model, using the depth-dependent ``Hillier'' clumping law \citep[e.g.,][]{Hillier+2003} with a characteristic velocity of $\varv_\text{cl} = 100\,\mathrm{km}\,\mathrm{s}^{-1}$. For the rest of the models, we assume a smooth wind ($D \equiv 1$), motivated by the low clumping factors and relatively smooth wind structure obtained in \citet{Moens+2022wr}.

The spectral synthesis is calculated by performing a radiative transfer calculation in the observer's frame on the converged models. 
In terms of elements, we assume the composition of a hydrogen-free WN, including He, C, N, O, Ne, Na, Mg, Al, Si, P, S, Cl, Ar, K, Ca, and the Fe group. For CNO, we have assumed the mass fractions of 98\% Helium, 1.5\% Nitrogen, 0.04\% Carbon and 0.1\% Oxygen \citep{Sander+2020}, and otherwise solar abundances from \citet{Asplund+2009}, while \citet{Moens+2022wr} used the set from \citet{GrevesseNoels1993}. We do not expect any significant impact of this abundance difference for any of our comparisons.

 \subsection{3D radiation-hydrodynamic framework}

The multi-dimensional framework solves the RHD partial differential equations (PDEs) on a `box-in-a-star' finite volume grid using the RHD module from \citet{Moens+2022fld} in the MPI-AMRVAC code \citep{Xia+2018}, including correction terms for spherical divergence \citep{Sundqvist+2018, Moens+2022fld}. While the 1D PoWR models preform a detailed non-LTE line opacity calculation \citep[with a superlevel approach for iron and iron group elements (Sc to Ni), see][]{Graefener+2002}, a hybrid approach by \citet{Poniatowski+2022} is used to compute the opacities instead: at the hydrostatic stellar core, the adopted tabulated Rosseland mean opacities from OPAL \citep{IglesiasRogers1996} are sufficient to lift up the gas, while in the optically thin supersonic regime the line opacities, enhanced due to Doppler shifts, are dominating. This enhancement is inferred from tabulated flux-weighted mean opacities \citep{Poniatowski+2022} calculated with the help of a CAK-like \citep*[named after][]{Castor+1975} force multiplier approach, modified by a line-strength cut-off introduced by \citet{Gayley1995} and accounting for space- and time-dependent line-force parameters.

In this work, we compare the three cWR star models by \citet{Moens+2022wr}, where the finite volume extends from the lower boundary inside the WR atmosphere at the (quasi-)hydrostatic core radius ($R_{c}=R_{\odot}$, located at $T_{c} \approx 261\,$kK, $289\,$kK, and $310\,$kK for the $\Gamma$2, $\Gamma$3, and $\Gamma$4 models, respectively) up to $6R_{c}$ the supersonic outflow region. We do not cover the lowest luminosity $\Gamma$1 model in \citet{Moens+2022wr} showing more of an O-type stellar wind and corresponding to a hot, stripped star (e.g., a He-dominated O subdwarf, but with higher luminosity than usually observed). A more extensive discussion of the main differences between the 1D stationary and the multi-dimensional RHD modelling approaches is given in \citet{GonzalezTora+25}.

 \section{Results and discussion}\label{sec:results}

      \begin{figure*}
      \centering
      \includegraphics[width=1.\linewidth]{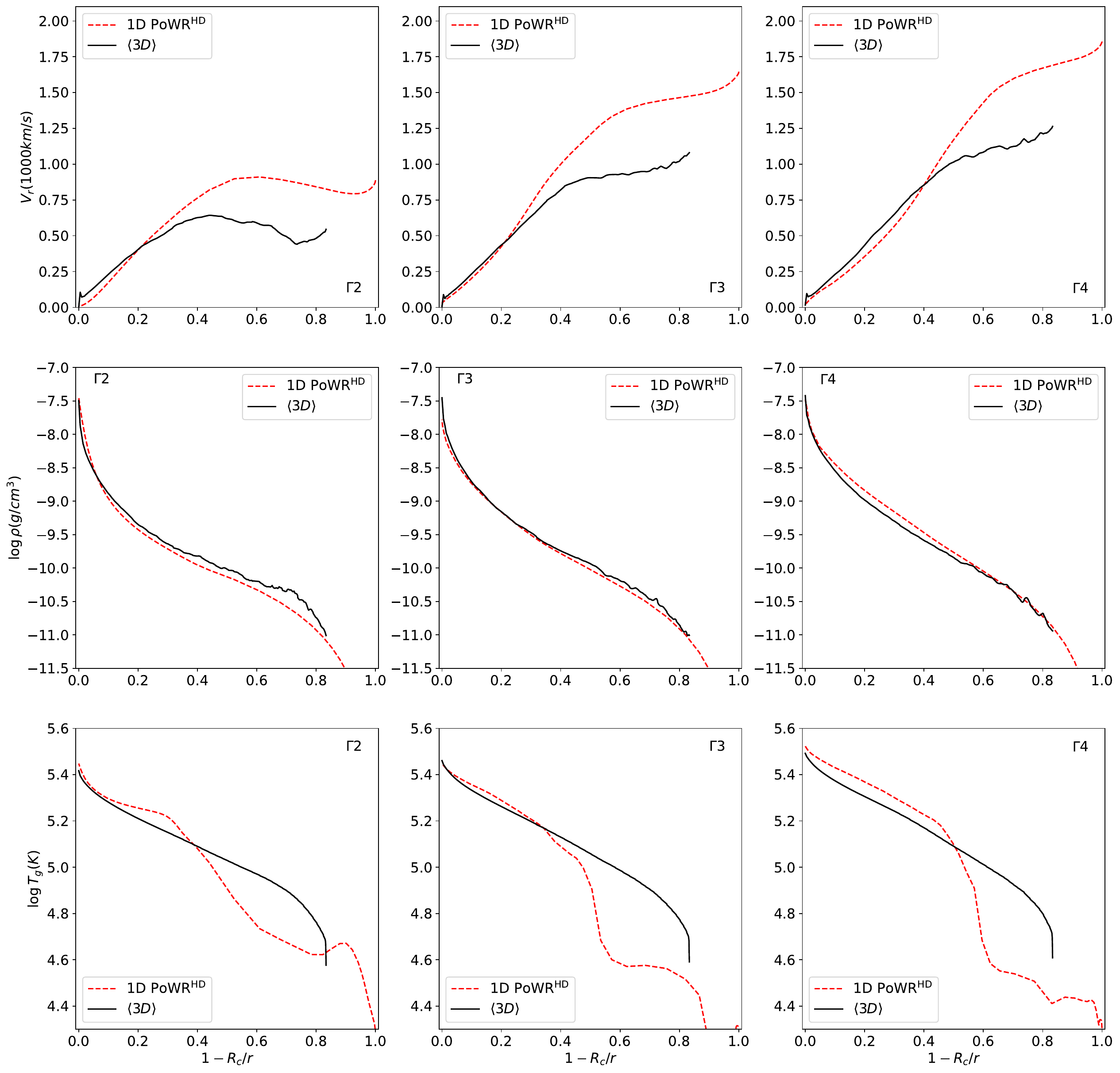}
      \caption{Profile comparison for the 3D averaged models (solid black) by \citet{Moens+2022wr} for the $\Gamma2$, $\Gamma3$ and $\Gamma4$ WR stars (from left to right) with 1D PoWR$^{\textsc{hd}}$ models using the same basic parameters (dashed-red). \textit{Upper panels:} Wind velocity profiles. \textit{Middle panels:}  density profiles \textit{Lower panels:} gas temperature profiles.}
      \label{fig:profile}
  \end{figure*}

        \begin{figure*}
      \centering
      \includegraphics[width=1.\linewidth]{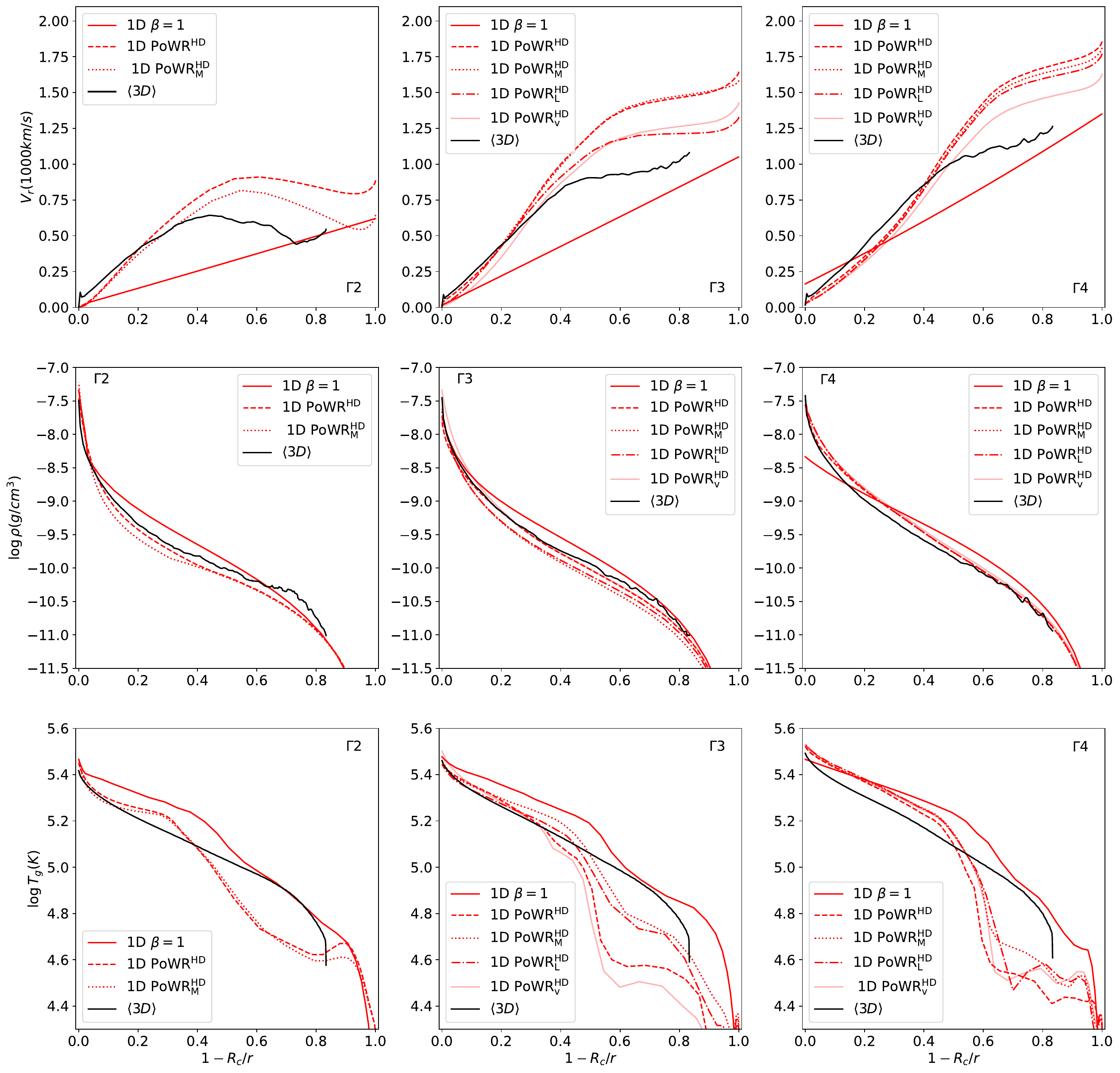}
      \caption{ Profile comparison for the $\Gamma2$, $\Gamma3$ and $\Gamma4$ stars. \textit{Upper panels: }Wind velocity profile, in solid black for the 3D averaged model of \citet{Moens+2022wr}, dashed-red for the 1D PoWR$^{\textsc{hd}}$ using the same basic parameters, dotted for the PoWR$^{\textsc{hd}}$ changing the mass, dash-dotted for the PoWR$^{\textsc{hd}}$ changing the luminosity value with respect to \citet{Moens+2022wr}, lighter-red for the PoWR$^{\textsc{hd}}$ with $\varv_{\mathrm{Dop}} = 50\,\mathrm{km}\,\mathrm{s}^{-1}$ and solid-red for the 1D PoWR approach with a $\beta$-law, from left to right for the $\Gamma2$, $\Gamma3$ and $\Gamma4$ WR stars, respectively. \textit{Middle panels: }Same as the \textit{upper panel} but for the density profile. \textit{Lower panel:} Same as the \textit{upper panel} but for the gas temperature. 
      }
      \label{fig:profile2}
  \end{figure*}

\subsection{3D vs 1D stratification comparisons}
We calculate both PoWR$^{\textsc{hd}}$ and $\beta$-law PoWR models to check how similar their converged velocity, density, and gas temperature stratifications are to the temporal averages from the density-weighted radial profile stratifications of the $\Gamma$2, $\Gamma$3, and $\Gamma$4 cWR 3D models from \citet{Moens+2022wr}. 

\subsubsection{Hydrodynamic 1D models with identical input}\label{sec:comp3D}

The first and most straight-forward approach is the direct comparison of the 3D results with 1D PoWR$^{\textsc{hd}}$ models using the same basic parameters as \citet{Moens+2022wr}, namely $R_\ast = R_\text{c} = 1\,R_\odot$, $M = 10\,M_\odot$ and the corresponding luminosities ($\log(L_{\star}/L_{\odot})$) of their $\Gamma 2$, $\Gamma 3$ and $\Gamma 4$ models. We have tuned the boundary Rosseland-mean optical depth accounting for the lines and the continuum, $\tau_{\text{max}}$, to have approximately the same value as for the 3D models, $\tau_{\text{max}}\sim30$, 35 and 40 for the $\Gamma 2$, $\Gamma 3$ and $\Gamma 4$ models, respectively. 
Fig.~\ref{fig:profile} shows in solid black the density-weighted radial velocities, the gas density and temperature profile stratifications for the spatially averaged 3D models from \citet{Moens+2022wr}\footnote{Additionally, the 3D averaged velocity profiles show an nonphysical spike at the lower boundary, this is due to numerical problems in the boundary conditions of the multi-dimensional simulations.} and the corresponding PoWR$^{\textsc{hd}}$ results in dashed-red.

As the mass-loss rates ($\dot{M}$) and terminal velocities ($\varv_{\infty}$) are iteratively determined in the PoWR$^{\textsc{hd}}$ models, they will not per se align with the values obtained by \citet{Moens+2022wr}, as we show in Table~\ref{tab:params}.
Indeed, the $\dot{M}$ and $\varv_{\infty}$ values of the 1D PoWR$^\textsc{hd}$ models are both higher than those of the 3D models from \citet{Moens+2022wr}. 
Yet, Fig.~\ref{fig:profile} shows that the density stratifications of the 1D models generally reproduce the overall behaviour of the 3D models. In particular the shapes of the velocity fields (upper panels in Fig.~\ref{fig:profile}) present a very similar trend to the 3D models. Noteably, they launch a bit further out and then always reach higher velocities before flattening or even decelerating in the case of $\Gamma$2 models. Note that the non-monotonic $\varv(r)$ for the 1D $\Gamma$2-equivalent is not the interpolated velocity field entering the CMF radiative transfer, but rather the last hydro solution from the PoWR$^\textsc{hd}$ model \citep[see][for more technical details on the handling of non-monotonic solutions]{Sander+2023}.

Moreover, while the 3D models extend up to $6\,R_\odot$, the 1D models are considerably more extended with their outer boundaries extending to $10\,000\,R_\odot$ or even $50\,000\,R_\odot$. Yet, as evident from Fig.~\ref{fig:profile} and Table~\ref{tab:params}, this further extension is not the main reason for the higher $\varv_{\infty}$ in the 1D models as they reach higher wind velocities already before surpassing the 'hot' iron bump which is the main reason for the flattening of $\varv(r)$. Here, the hot iron bump refers to the opacity peak due to the recombination of iron at $T\sim$$200\,$kK, hotter than the 'cool' iron bump for WR stars, typically between $T\sim$$35-70\,$kK instead \citep[e.g.,][]{NugisLamers2002,GraefenerHamann2008}.

The lower $\dot{M}$ in \citet{Moens+2022wr} compared to the 1D could have several explanations. On the one hand, because of the mismatch in the methodologies when estimating the profile stratifications: as mentioned in \citet{GonzalezTora+25}, the average velocity profile $\langle\varv\rangle$ will be greater than the density averaged velocity, $\langle\rho\varv\rangle/\langle\rho\rangle$ in the multi-D models \citep[e.g.,][]{Moens+2022wr,Debnath+2024}. Therefore, a higher $\dot{M}$ results when matching the same $\langle\rho\rangle$. On the other hand, due to the variability of the $\dot{M}$ and luminosity both in space and time during the 3D simulations, there is also an uncertainty on the resulting 3D averages which we treat as ``goal'' values in this work. As notable in Fig. 4 of \citet{Moens+2022wr}, we can expect a factor of at least 0.2 dex both in the $\dot{M}$ and luminosity dispersion from the mass flux during the time-dependent simulations. Our $\lesssim$0.2 dex difference in $\dot{M}$ between 1D and 3D modelling approaches is below this dispersion of the 3D simulations, meaning that even the standard setup of the 1D models yields a very good agreement. Still, different PoWR$^{\textsc{hd}}$ input adjustments will be tested in Sect.~\ref{sec:powrs} to match the 3D $\dot{M}$ values even better.

The density stratifications for the 1D and 3D models match well (middle panels in Fig.~\ref{fig:profile}).  
At the same radial inner boundary ($R_{c}=1\,R_{\odot}$), the 1D models obtain inner-most temperatures of $T_{c} \approx 281$, 282 and 334\,kK (or $\log T_{c}=5.44$, 5.45, 5.52) for the $\Gamma$2, $\Gamma$3 and $\Gamma$4 models, respectively. 
The corresponding 3D models have $T_{c} \approx 261$, $289$ and 310\,kK (equivalent to $\log T_{c}=5.42$, 5.46 and 5.49) for the $\Gamma$2, $\Gamma$3 and $\Gamma$4 models, respectively. Outwards, both 1D and 3D models follow a similar temperature trend until the start of the outer wind layers ($r\sim 1.6\,R_{\mathrm{c}}$), where the 1D models show a strong temperature decrease with a dip reaching its minimum at $r\sim 2.5-3\,R_{\odot}$. As discussed in \citet{GonzalezTora+25}, the differences in temperature profiles are likely due to the different methods used to compute the energy balance in the 1D and the 3D approaches. In contrast to the non-LTE regime for 1D models, the 3D models approximated the energy and Planck mean opacities by the flux mean, forcing the gas and radiation temperatures to be the same, essentially acting as an LTE regime with the net effect being a higher gas temperature \citep[see][for a detailed discussion]{GonzalezTora+25}. 
The radiation temperature calculated from the integrated mean intensity as well as the flux temperature calculated from the total emergent flux for the PoWR model is shown in Fig.~\ref{fig:trad}.

 \begin{table*}
\caption{Comparison of the resulting wind parameters and effective temperatures predicted by the 1D PoWR$^\textsc{hd}$ and the 3D RHD models from \citet[denoted as M+22]{Moens+2022wr} when using the same basic input parameters. }
\label{tab:params}
\small
\centering
\begin{tabular}{c c c c c c c c c c}
    \hline \hline
       Model & $\log(L_{\star}/L_{\odot})$ & $\log\dot{M}$ & $\log\dot{M}^{M+22}$  & $\varv_{\infty}$ & $\varv(6\,R_\ast)$  & $\varv^{M+22}_{\infty}$  & $\log T_{\mathrm{eff}}$  & $\log T_{\mathrm{eff}}^{M+22}$ & $M_{\star}/M_{\odot}$  \\
        &  & ($M_{\odot}\text{yr}^{-1}$) & ($M_{\odot}\text{yr}^{-1}$) & ($10^3$ $\text{km s}^{-1}$) & ($10^3$ $\text{km s}^{-1}$) & ($10^3$ $\text{km s}^{-1}$) & (K) & (K) &  \\
     \hline 
         $\Gamma 2$ & 5.47  & -4.64 & -4.82  & 0.88 & 0.83 & 0.62 & 4.69 & 4.83 & 10 \\
         $\Gamma 3$ & 5.64  & -4.36 & -4.49  & 1.65 & 1.65 & 1.05 & 4.72 & 4.89 & 10 \\
         $\Gamma 4$ & 5.74  & -4.11 & -4.14 & 1.86 & 1.85 & 1.35 & 4.66 & 4.88 & 10 \\
     \hline
     
\end{tabular}
\tablefoot{The effective temperatures are given for $\tau_\text{Ross}=2/3$.}
\end{table*}

\subsubsection{Hydrodynamic 1D models with modified input}\label{sec:powrs}

Given the differences in the obtained $\dot{M}$ in Sect.\,\ref{sec:comp3D}, we explore how much the input of 1D PoWR$^{\textsc{hd}}$ models would need to be adjusted in order to yield the same $\dot{M}$ as the 3D models from \citet{Moens+2022wr}. Since the density profile is already well matched between both methodologies, obtaining the same $\dot{M}$ would give a better velocity profile agreement. For this purpose, we use the option in PoWR$^{\textsc{hd}}$ to keep $\dot{M}$ fixed, but instead change either the stellar luminosity $L$ or the stellar mass $M$ during the hydrodynamic stratification updates. Given that the 1D models obtain systematically higher mass-loss rates, we expect that either a reduction of $L$ or an increase in $M$ is necessary to reach the 3D $\dot{M}$ values. As discussed above, the luminosity of a 3D model is not a global quantity and thus subject to inherent fluctuations which imply that different assumptions could be used in a 1D approach. The luminosity-scaled models thus give us an indication whether necessary $L$ adjustments to reproduce the 3D mass-loss rate are within the observed 3D fluctuations. The mass-scaled models instead give us an indication for forthcoming spectral analysis efforts of WR stars. For WR stars that do not have an orbital mass, dynamically-consistent models can provide a unique handle on the stellar mass. By determining the difference in $M$ between the inherent 1D solution and the mass adjustments to reproduce the 3D mass-loss rate, our test calculations here provide us with a first uncertainty estimate for forthcoming derived masses from spectral analysis with dynamically-consistent models.

In Fig.~\ref{fig:profile2}, we show the same curves as in Fig.\,\ref{fig:profile} adding the different PoWR$^{\textsc{hd}}$ solutions with modified $L$ or $M$ where $\dot{M}$ is fixed to the 3D value: For the $\Gamma 2$ and $\Gamma 3$ case, the 1D model changes the stellar masses from initially 10$M_{\odot}$ to 11.3$M_{\odot}$ (dotted profile, PoWR$^{\textsc{hd}}_{\mathrm{M}}$). For $\Gamma 3$, the luminosity decreases to $\log(L_{\star}/L_{\odot})=5.56$ (dash-dotted profile,  PoWR$^{\textsc{hd}}_{\mathrm{L}}$), while no solution with decreased luminosity could be found for $\Gamma 2$. In the case of $\Gamma 4$, a change in the stellar mass to 10.2$M_{\odot}$ (dotted profile, PoWR$^{\textsc{hd}}_{\mathrm{M}}$) or a luminosity change to $\log(L_{\star}/L_{\odot})=5.72$ (dash-dotted profile, PoWR$^{\textsc{hd}}_{\mathrm{L}}$) is needed to obtain the same $\dot{M}$ as \citet{Moens+2022wr}.  We have also included in Fig.~\ref{fig:profile2} the PoWR$^{\textsc{hd}}$ solution with a lower Doppler velocity (light-red profiles, PoWR$^{\textsc{hd}}_{\mathrm{v}}$) discussed in Sect.~\ref{sec:vdop}.  All modified stellar parameters are shown in Table~\ref{tab:paramsbeta} for an easier comparison. 

  \begin{table*}
\caption{Stellar parameters for the $\beta$-law and modified hydrodynamical PoWR models to match the profile of \citet{Moens+2022wr}.  
}
\label{tab:paramsbeta}
\small
\centering
\begin{tabular}{c c c c c c c c c}
    \hline \hline
       Model & $\log(L_{\star}/L_{\odot})$ & $\log(\dot{M}/M_{\odot}\text{yr}^{-1})$ & $\varv_{\infty}$ ($10^3$ $\text{km s}^{-1}$) & $\log T_{\mathrm{eff}}$ (K)  & $M_{\star}/M_{\odot}$ & $\varv_{\mathrm{Dop}}$ ($\text{km s}^{-1}$) & Framework \\
     \hline 
         $\Gamma 2$ & 5.47 & -4.82 & 0.62 & 4.70 & 10 & 100 & $\beta=1$\\
         $\Gamma 2$ & 5.47 & -4.82 & 0.65 & 4.70 & 11.3 & 100 & PoWR$^{\textsc{HD}}_\text{M}$\\
    \hline 
         $\Gamma 3$ & 5.64 & -4.49 & 1.05 & 4.70 & 10 & 100 & $\beta=1$\\
         $\Gamma 3$ & 5.64 & -4.50 & 1.58 & 4.81 & 11.3 & 100 & PoWR$^{\textsc{HD}}_\text{M}$\\
         $\Gamma 3$ & 5.56 & -4.50 & 1.32 & 4.73 & 10 & 100 & PoWR$^{\textsc{HD}}_\text{L}$\\
         $\Gamma 3$ & 5.64 & -4.41 & 1.43 & 4.74 & 10 & 50 & PoWR$^{\textsc{HD}}_\text{v}$\\
    \hline 
         $\Gamma 4$ & 5.74 & -4.14 & 1.35 & 4.61 & 10 & 100 & $\beta=1$\\
         $\Gamma 4$ & 5.74 & -4.12 & 1.82 & 4.66 & 10.2 & 100 & PoWR$^{\textsc{HD}}_\text{M}$\\
         $\Gamma 4$ & 5.72 & -4.12 & 1.77 & 4.65 & 10 & 100 & PoWR$^{\textsc{HD}}_\text{L}$\\
         $\Gamma 4$ & 5.74 & -4.13 & 1.63 & 4.65 & 10 & 50 & PoWR$^{\textsc{HD}}_\text{v}$\\

     \hline
     
\end{tabular}
\tablefoot{All fundamental parameters are scaled at $\tau=2/3$. All the models have a $\varv_{\text{turb}}=21\,\mathrm{km}\,\mathrm{s}^{-1}$.}
\end{table*}

The obtained differences in $\log L/L\odot$ of $0.08$ ($\Gamma 3$) and $0.02$ ($\Gamma 4$) are well below the luminosity dispersion of the 3D models. The required changes in $M$ are also very moderate and typically below the accuracy reachable in orbital mass determinations for WR stars \citep[e.g.,][]{Richardson+2021}. In most cases, the adjusted PoWR$^{\textsc{HD}}$ models show no significant changes to the density and temperature profiles compared to the initial solutions (see Fig.~\ref{fig:profile2}). Concerning the radial-directed velocities, the launch of the wind is consistently further out when comparing to the 3D density-averaged velocities. The flattening still occurs at higher velocities than for the 3D models and also the outermost velocities (even when comparing at the outer limit $6\,R_\ast$ of the 3D models) are notably higher, with the exception of the mass-adjusted $\Gamma 2$-model. 

\subsubsection{Hydrodynamic 1D models with turbulent pressure, different Doppler velocities and clumping factors}\label{sec:vdop}

One parameter that usually plays a more technical role in the calculation of 1D atmosphere models is the Doppler velocity ($\varv_{\mathrm{Dop}}$) used in the CMF radiative transfer. In contrast to the spectral synthesis, where this value is usually calculated directly from the thermal and microturbulent broadening, the CMF calculations use a fixed, depth-independent value which not only defines the resolution of the frequency grid, but also enters the Gaussian profiles assumed for the opacity calculations. For WR models, this value is typically set to $\varv_{\mathrm{Dop}} = 100\,\mathrm{km}\,\mathrm{s}^{-1}$, accounting for both thermal and turbulent broadening. While $\varv_{\mathrm{Dop}}$ as such does not enter the hydrodynamic equation, it indirectly impacts the radiative force calculated in the CMF, in particular in the subsonic regime where a larger value can lead to a slightly higher $a_\text{rad}$ due to broadening the otherwise very narrow profiles.

In order to obtain wind parameters and structure profiles closer to the models from \citet{Moens+2022wr}, we also investigate the effect on changing the Doppler velocities ($\varv_{\mathrm{Dop}}$) in the PoWR$^\textsc{hd}$ models. The stellar parameters as well as the profile stratifications are shown in Table~\ref{tab:paramsbeta} and Fig.~\ref{fig:profile2} in light-red for the $\Gamma3$ and $\Gamma4$ models. For the $\Gamma2$ representation, the $\log\,\dot{M}$ values dropped below $-5.0$, where we stopped the calculations as the 3D model yields $-4.82$. In the $\Gamma$2 case, the opacities of the hot iron bump seem to be too low when just assuming $\varv_{\mathrm{Dop}} = 50\,\mathrm{km}\,\mathrm{s}^{-1}$ and our default turbulent velocity of $21\,\mathrm{km}\,\mathrm{s}^{-1}$. For the $\Gamma3$-equivalent 1D model, a simple reduction to $\varv_{\mathrm{Dop}} = 50\,\mathrm{km}\,\mathrm{s}^{-1}$ indeed reduces the discrepancy in $\dot{M}$ obtained, namely from $-4.36$ to $-4.41$, which is still $0.08$\,dex above the 3D result.  Interestingly, $\varv_{\mathrm{Dop}} = 50\,\mathrm{km}\,\mathrm{s}^{-1}$ seems to be a good choice in the $\Gamma4$ case, where the reduction from $100\,\mathrm{km}\,\mathrm{s}^{-1}$ to $50\,\mathrm{km}\,\mathrm{s}^{-1}$ changes the mass-loss rate from $-4.11$ to $-4.13$, essentially coinciding with the 3D value of $-4.14$. Yet, this agreement does not hold for the terminal velocities, although the value of $1630\,\mathrm{km}\,\mathrm{s}^{-1}$ is closer to the $1350\,\mathrm{km}\,\mathrm{s}^{-1}$ result from the 3D simulations. 
  
Our test calculations reveal that there seems to be no fixed value of $\varv_{\mathrm{Dop}}$ that could generally lead to a better alignment between 1D and 3D results. Interestingly, the different results actually show that higher values of $\varv_{\mathrm{Dop}}$ are required for models with lower mass-loss rates, further away from the Eddington Limit. Based on a further test calculation with an even lower $\varv_{\mathrm{Dop}} = 20\,\mathrm{km\,s^{-1}}$ for the $\Gamma 3$-model, we obtain a scaling for $\log\dot{M}$ with $\log\varv_{\mathrm{Dop}}$:
\begin{equation}\label{eq:slope}
    \frac{\partial\log\dot{M}}{\partial\log\varv_{\mathrm{Dop}}}\simeq0.26\text{,}
\end{equation}
as illustrated in Fig.~\ref{fig:slope}. Note that both the slope and value differ from the microturbulent velocity scaling found for OB star models by \citet{Bjoerklund+2021}. For the WR stars, we assume higher $\varv_{\mathrm{Dop}}$ values than in the OB star regime since the stellar lines are very broad and a high $\varv_{\mathrm{Dop}}$ minimizes the computational time with usually no noticeable effect on the spectral synthesis in case of optically thick winds. 

All PoWR$^{\textsc{HD}}$ models presented up to this point include a small turbulent pressure in the hydrodynamic equation with a turbulent velocity of value of $21\,\mathrm{km}\,\mathrm{s}^{-1}$. When comparing to the O supergiant findings by \citet{Debnath+2024}, this value might be too low, in particular in the regime of the $\Gamma 2$-model where the lower $\varv_{\mathrm{Dop}}$-model did not converge and the opacity increase from the higher $\varv_{\mathrm{Dop}}$ might indirectly counter the effect of an underestimated turbulent pressure.

\begin{figure}
      \centering
      \includegraphics[width=.85\linewidth]{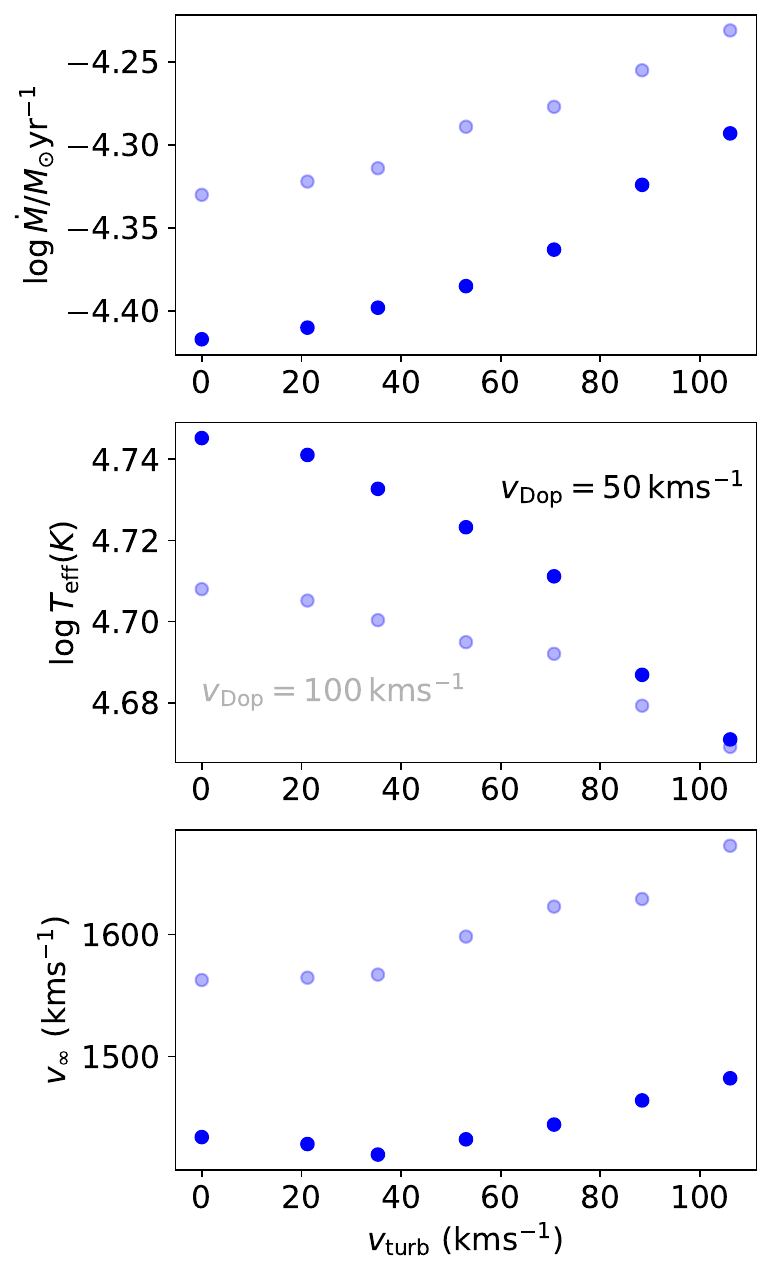}
      \caption{ Different PoWR$^{\textsc{hd}}$ models for $\Gamma$3 with $\varv_{\mathrm{Dop}}=100\,\mathrm{km}\,\mathrm{s}^{-1}$ (light color) and $\varv_{\mathrm{Dop}}=50\,\mathrm{km}\,\mathrm{s}^{-1}$ (dark color) including a turbulence $\varv_{\mathrm{turb}}$ with respect to the $\log\dot{M}/M_{\odot}\text{yr}^{-1}$, $\log T_{\mathrm{eff}}$ and $\varv_{\infty}$.
      }
      \label{fig:vturbmdot}
  \end{figure}

\begin{figure}
      \centering
      \includegraphics[width=.9\linewidth]{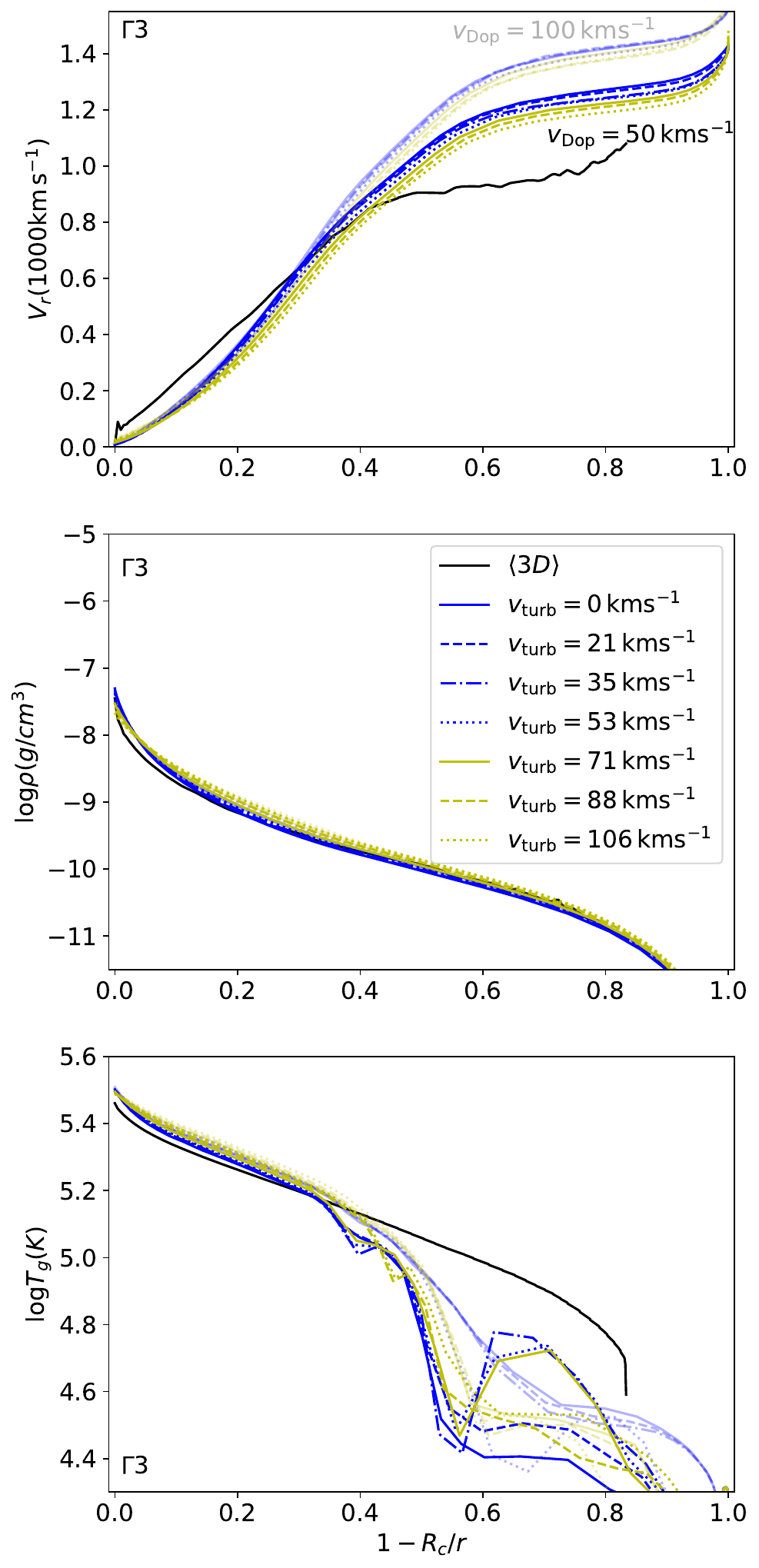}
      \caption{ Radial velocity, density and gas temperature profiles for the averaged 3D model from \citet{Moens+2022wr} in solid black and  PoWR$^{\textsc{hd}}$ models for $\Gamma$3 $\varv_{\mathrm{Dop}}=100\,\mathrm{km}\,\mathrm{s}^{-1}$ (light color) and $\varv_{\mathrm{Dop}}=50\,\mathrm{km}\,\mathrm{s}^{-1}$ (dark color) including different turbulence values $\varv_{\mathrm{turb}}$.
      }
      \label{fig:profilevturb}
  \end{figure}

To investigate the combined effect of $\varv_{\mathrm{Dop}}$ and turbulent pressure, we now vary the latter component in Eq.\,\eqref{eq:apress} by changing $\varv_{\mathrm{turb}}$, similarly to \citet{GonzalezTora+25}. 
Figure~\ref{fig:vturbmdot} shows the $\log\dot{M}/M_{\odot}\,\text{yr}^{-1}$, $\log T_{\mathrm{eff}}$, and $\varv_{\infty}$ obtained when including different (radially constant) $\varv_{\text{turb}}$ values ranging from $0$ to $106\,\mathrm{km}\,\mathrm{s}^{-1}$ with $\Delta\varv_{\mathrm{turb}}\sim20\,\mathrm{km}\,\mathrm{s}^{-1}$ for $\Gamma$3 assuming a Doppler velocity of $\varv_{\mathrm{Dop}}=100\,\mathrm{km}\,\mathrm{s}^{-1}$ (light color) and a lower  $\varv_{\mathrm{Dop}}=50\,\mathrm{km}\,\mathrm{s}^{-1}$ (dark color). The velocity, density, and temperature stratifications for the $\Gamma$3 PoWR$^{\textsc{hd}}$ models with different $\varv_{\mathrm{turb}}$ are shown in Fig.~\ref{fig:profilevturb} with respect to the averaged 3D profiles for the $\Gamma$3 model from \citet{Moens+2022wr} for $\varv_{\mathrm{Dop}}=100\,\mathrm{km}\,\mathrm{s}^{-1}$ (light color) and $\varv_{\mathrm{Dop}}=50\,\mathrm{km}\,\mathrm{s}^{-1}$ (dark color).

As shown in Fig.~\ref{fig:profilevturb}, the default turbulent velocity of $21\,\mathrm{km}\,\mathrm{s}^{-1}$ has a very small effect on the derived wind parameters and therefore cannot explain the higher $\dot{M}$ in comparison to the 3D models. However, when lowering the $\varv_{\mathrm{Dop}}$, we obtain overall reduced values of the global stellar parameters and profiles for the same value of $\varv_{\mathrm{turb}}$, getting closer to the results from the averaged 3D models.

\begin{figure}
      \centering
      \includegraphics[width=.85\linewidth]{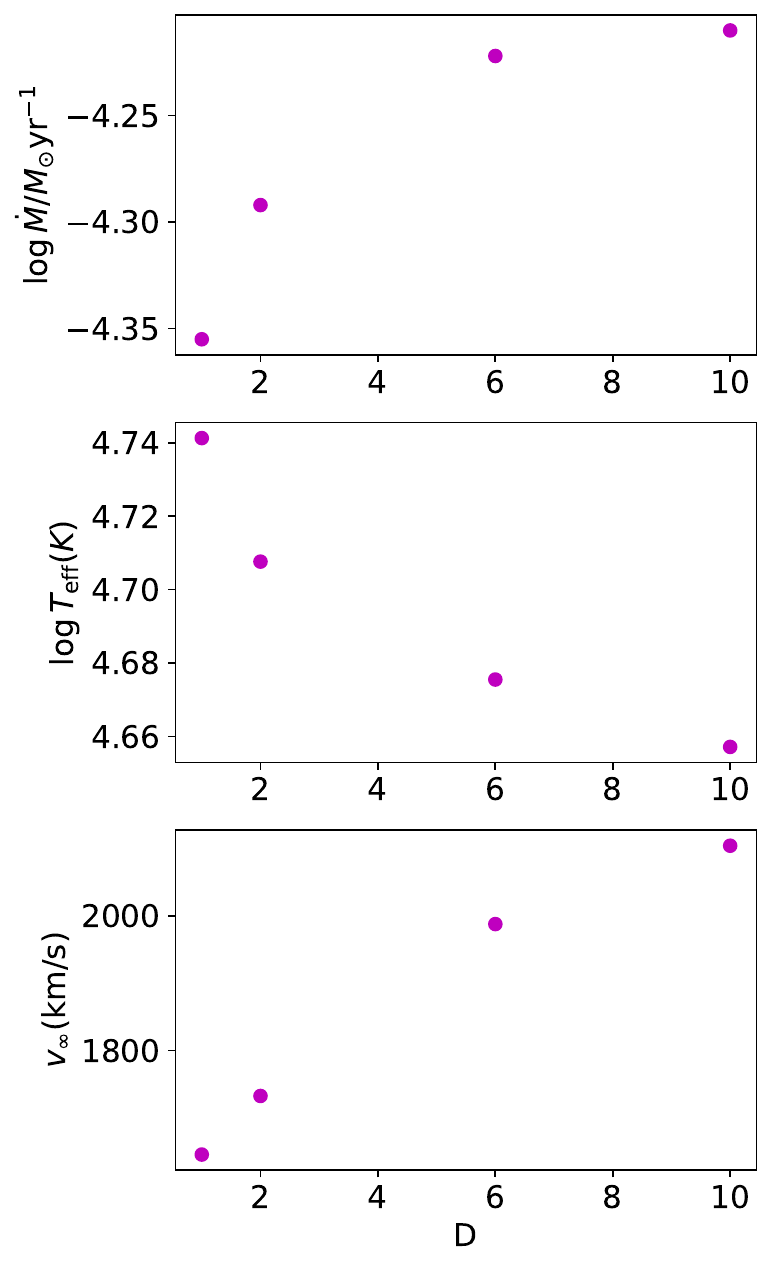}
      \caption{ Different PoWR$^{\textsc{hd}}$ models for $\Gamma$3 including a turbulence $\varv_{\mathrm{turb}}=21\,\mathrm{km}\,\mathrm{s}^{-1}$ with respect to the $\log\dot{M}/M_{\odot}\text{yr}^{-1}$, $\log T_{\mathrm{eff}}$ and $\varv_{\infty}$.
      }
      \label{fig:densconmdot}
  \end{figure}

\begin{figure}
      \centering
      \includegraphics[width=.9\linewidth]{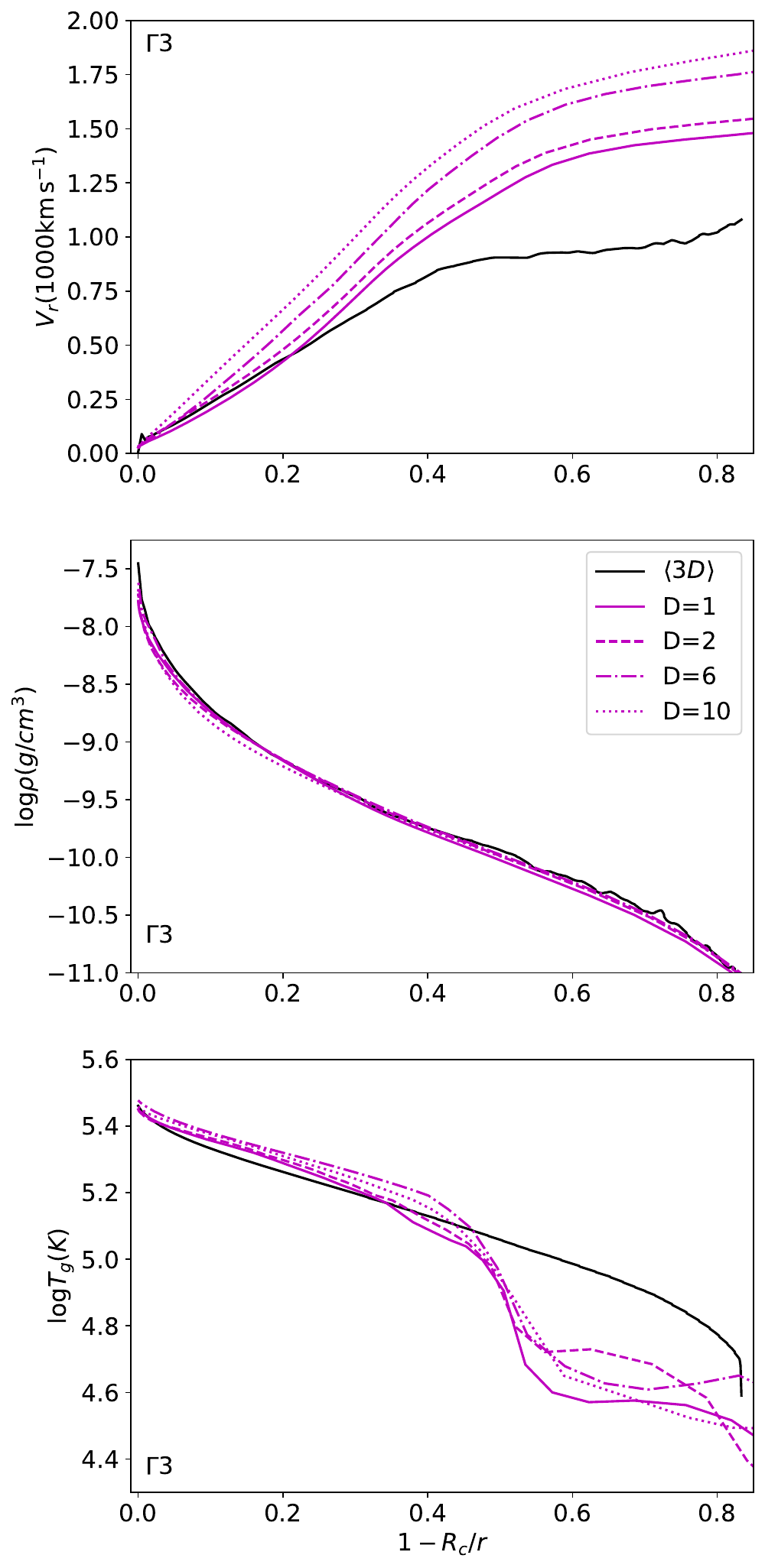}
      \caption{ Radial velocity, density and gas temperature profiles for the averaged 3D model from \citet{Moens+2022wr} in solid black and  PoWR$^{\textsc{hd}}$ models for $\Gamma$3 including different turbulence values $\varv_{\mathrm{turb}}$.
      }
      \label{fig:profiledenscon}
  \end{figure}

In addition, we explore the effect on changing the density contrast, $D$, in the PoWR models. $D$ denotes the overdensity factor of the clumps compared to a smooth wind. The formal definitions vary a bit between different atmosphere codes \citep[see][for a recent comparison]{Sander+2024}, but in the limit of optically thin clumping with a void interclump medium, as used in this work, $D$ is identical to $f_\text{cl}$ and the inverse of the volume filling factor. Fig.~\ref{fig:densconmdot} shows the output $\log\dot{M}/M_{\odot}\text{yr}^{-1}$, $\log T_{\mathrm{eff}}$ and $\varv_{\infty}$ with respect to $D$ for a $\Gamma$3 PoWR$^{\textsc{hd}}$ model with the same stellar parameters as Table~\ref{tab:params} and $\varv_{\mathrm{Dop}}=100\,\mathrm{km}\,\mathrm{s}^{-1}$. The clumping increases from smooth inner with following the clumping law from \citet{Hillier+2003} with a characteristic velocity of 100 km\,s$^{-1}$. Fig.~\ref{fig:profiledenscon} shows the radial velocity, density and gas temperature stratifications for 1D models with different $D$ parameters with respect to the 3D averaged $\Gamma$3 model from \citet{Moens+2022wr}. 

The upper panels in Figs.~\ref{fig:vturbmdot} and \ref{fig:densconmdot} show that we obtain a higher $\log\dot{M}/M_{\odot}\text{yr}^{-1}$ when increasing the $\varv_{\mathrm{turb}}$ or $D$. 
Even with the lowest assumptions of a smooth model ($D=1$) and no turbulent pressure ($\varv_{\mathrm{turb}}=0\,\mathrm{km}\,\mathrm{s}^{-1}$) we still cannot reach the low $\log\dot{M}/M_{\odot}\text{yr}^{-1}$ values predicted by the 3D models for $\Gamma3$. The middle panels in  Figs.~\ref{fig:vturbmdot} and \ref{fig:densconmdot} show an anti-correlation between increased $\varv_{\mathrm{turb}}$ or $D$ and lower $\log T_{\mathrm{eff}}$, which is a direct consequence of the corresponding change in $\dot{M}$.
In all cases, the $T_\text{eff}$ value from \citet{Moens+2022wr} is again underestimated by $\sim$$30$\,kK. However, the different calculation methods of $T_{\mathrm{eff}}$ for the multi-dimensional simulations present a high dispersion (e.g., already to the order of $\sim$kK for O stars). In addition, the different temperature estimation methods between the 1D and 3D methodologies mentioned above (i.e., the 3D models forcing the gas and radiation temperatures to be the same) can also help explain the mismatch between both $T_{\mathrm{eff}}$ values. 

The lower panels in Figs.~\ref{fig:vturbmdot} and \ref{fig:densconmdot} show the terminal velocity ($\varv_{\infty}$). While the effect of increasing the turbulence is small, presenting just a difference of $50\,\mathrm{km}\,\mathrm{s}^{-1}$ for an increase of $\Delta\varv_{\mathrm{turb}}=60\,\mathrm{km}\,\mathrm{s}^{-1}$, increasing the density contrast has a strong effect on $\varv_{\infty}$ with a difference of almost $500\,\mathrm{km}\,\mathrm{s}^{-1}$ between a smooth model and a clumped model with $D=10$. This effect is also visible in the velocity profiles of Figs.~\ref{fig:profilevturb} and \ref{fig:profiledenscon}, while the density and temperature profiles do not present significant $\varv_{\mathrm{turb}}$ or $D$ dependency.

\subsubsection{1D $\beta$-law models}\label{sub:beta}

Fig.~\ref{fig:profile2} also shows the wind velocity, density and gas temperature profiles from the $\beta$-law PoWR approach using the solution of the hydrostatic equation in the (quasi-)hydrostatic regime  and a $\beta$-law (Eq.~\ref{eq:vbeta}) for the supersonic wind regime, commonly adopting $\beta=1.0$ for WR stars \citep{Hillier1988,Hamann+1988,HillierMiller1999} in solid red. Table~\ref{tab:paramsbeta} shows the final parameters for the $\beta$-law approach with the same $R_{\star}$, $\log(L_{\star}/L_{\odot})$,  $M_{\star}/M_{\odot}$, $\varv_\infty$ and $\log\dot{M}$ as \citet{Moens+2022wr} for the $\Gamma$2, $\Gamma$3 and $\Gamma$4 models, respectively. 

The main difference between the $\beta$-law and the hydrodynamically consistent 1D modelling approaches illustrated in Fig.~\ref{fig:profile2} is the wind velocity in the outer layers: while the hydrodynamical solution shows a flattening or even deceleration at $1-R_{\text{c}}/r\sim0.45$, the $\beta$-law approach has a continuing velocity increase and cannot account for any deceleration in the velocity due to the dense medium in the outer layers of WR stars. Hence, it predicts an unrealistic profile, which will not change even if $\varv_\infty$ would be adjusted to the 3D results. Specifically for the $\Gamma 4$-model, the $\beta$-law model starts already in the supersonic regime with a non-zero radial velocity. 
The $\beta$-law model still give a reasonable reproduction of the general density structure (cf.\ middle panels in  Fig.\,\ref{fig:profile2}), although considerable deviations are also noticeable for the inner part of the $\Gamma 4$-model. The temperature profiles show the usual deviations in the outer wind regime with the $\beta$-law model being significantly hotter in the inner wind. However, as we will discuss in Sect.~\ref{sec:spectra}, the different temperature profile do not significantly affect the resulting spectral line predictions.

\subsection{Spectral synthesis for PoWR models}\label{sec:spectra}
\begin{figure}
      \centering
      \includegraphics[width=1.\linewidth]{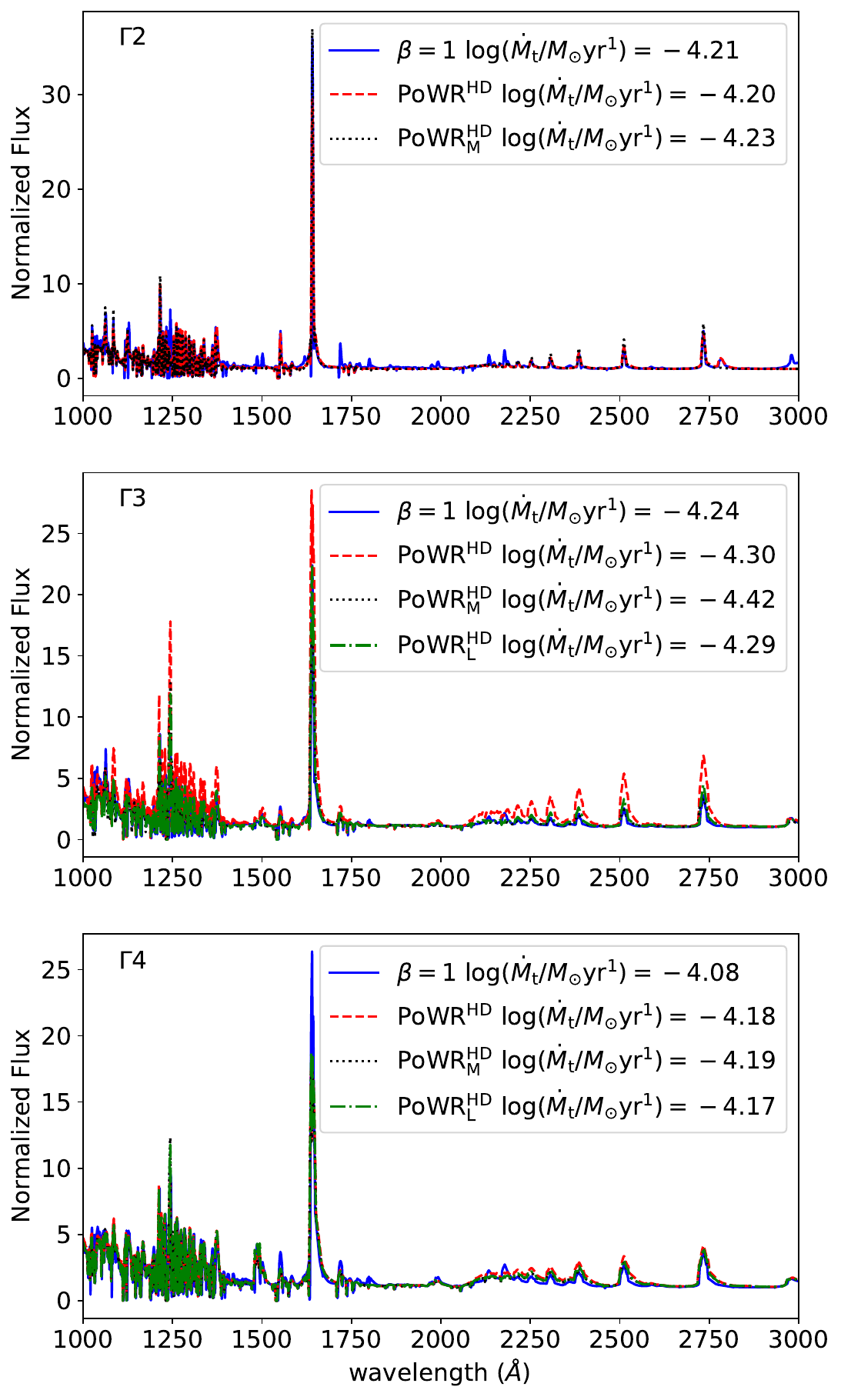}
      \caption{ Computed UV spectra for the PoWR models. \textit{From up to down:} the $\beta$-law approach (blue solid line), the PoWR$^{\textsc{hd}}$ with identical input as \citet{Moens+2022wr} (red dashed dotted), the PoWR$^{\textsc{hd}}$ with flexible mass (red dotted line) and the PoWR$^{\textsc{hd}}$ with flexible luminosity (red dash-dotted line), for $\Gamma$2, $\Gamma$3 and $\Gamma$4 models, respectively.
      }
      \label{fig:iue}
  \end{figure}

\begin{figure}
      \centering
      \includegraphics[width=1.\linewidth]{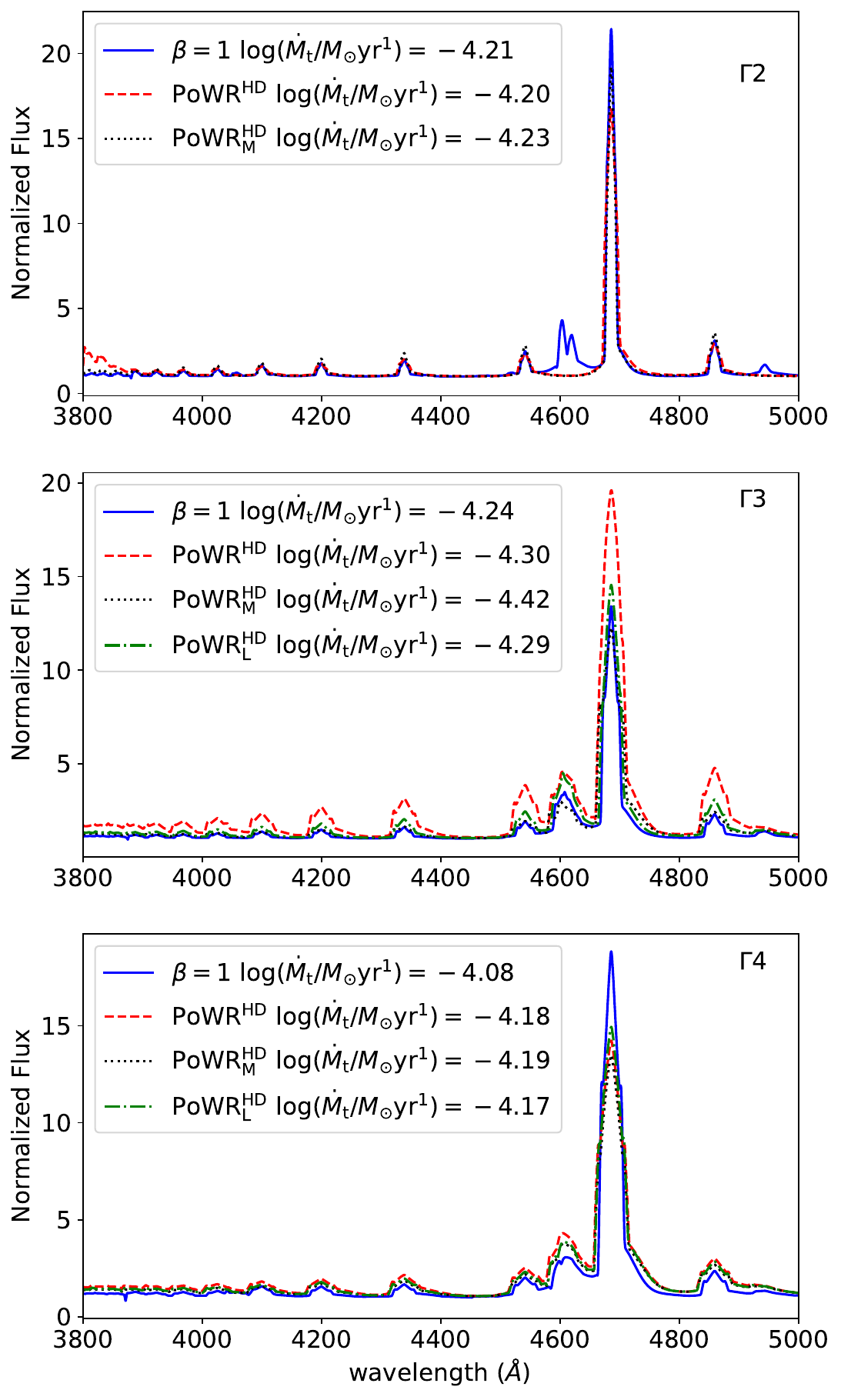}
      \caption{ Same as Fig.~\ref{fig:iue} but for the Optical spectral region.
      }
      \label{fig:opt}
  \end{figure}

To study the implications of the stratifications predicted by the 3D calculations from \citet{Moens+2022wr}, we compared the resulting synthetic spectra from the different 1D PoWR approaches. 
Figures~\ref{fig:iue} and \ref{fig:opt} show the computed UV and optical spectra for the $\beta$-law PoWR framework (solid blue), and the different PoWR$^{\textsc{hd}}$ approaches: the initial models with fixed stellar mass and luminosity (dashed red), the model with flexible stellar mass (dotted black) and with flexible luminosity (dash-dotted green) for the $\Gamma$2, $\Gamma$3 and $\Gamma$4 models, respectively. We also include in the legend the so-called transformed mass loss rate, $\dot{M}_{\mathrm{t}}$, defined as:
\begin{equation}
   \dot{M}_{\mathrm{t}}=\dot{M}\sqrt{D} \left( \frac{1000\mathrm{km s}^{-1}}{\varv_{\infty}} \right)\left( \frac{10^{6}L_{\odot}}{L} \right)^{3/4}\mathrm{.}
\end{equation}
This quantity was introduced by \citet{GraefenerVink2013} and refers to the inferred mass-loss rate the star would have with a $D=1$, luminosity of $10^{6}\,L_{\odot}$ and a $\varv_{\infty}=1000\,\mathrm{km}\,\mathrm{s}^{-1}$. Similar to the context of the ``transformed radius'' \citep{Schmutz+1989}, stars with a similar $\dot{M}_{\mathrm{t}}$ should have a similar spectrum. Checking Figs.\,\ref{fig:iue} and \ref{fig:opt} underlines that this is indeed the case. A zoomed-in version for the UV spectra is shown in Figure~\ref{fig:iue2}.

In the case of $\Gamma$2, the main difference between $\beta$-law and PoWR$^{\textsc{hd}}$ models is in the \ion{N}{v}$\,4604\,$\AA\, and \ion{N}{iii}$\,4640\,$\AA\, optical lines. This is due to the different ionization states for the $\beta$-law and hydrodynamical models: In the outer layers of the hydrodynamical models the \ion{He}{iii} has recombined to \ion{He}{ii}, as well as \ion{N}{v} to \ion{N}{iv}. This is not the case for the $\beta$-law model. 
For $\Gamma$3, both UV and Optical present overall higher emission lines for the initial PoWR$^\textsc{hd}$ model. This difference can be explained with the $\log\dot{M}/M_{\odot}\,\mathrm{yr}^{-1}=-4.36$ which is $\sim$0.15 dex higher for the PoWR$^\textsc{hd}$ model when compared to all the other approaches.  
For $\Gamma$4, all model approaches present a very similar UV and optical spectra, with very close agreement between the stellar parameters in Tab.~\ref{tab:paramsbeta}.

\begin{figure}
      \centering
      \includegraphics[width=1.\linewidth]{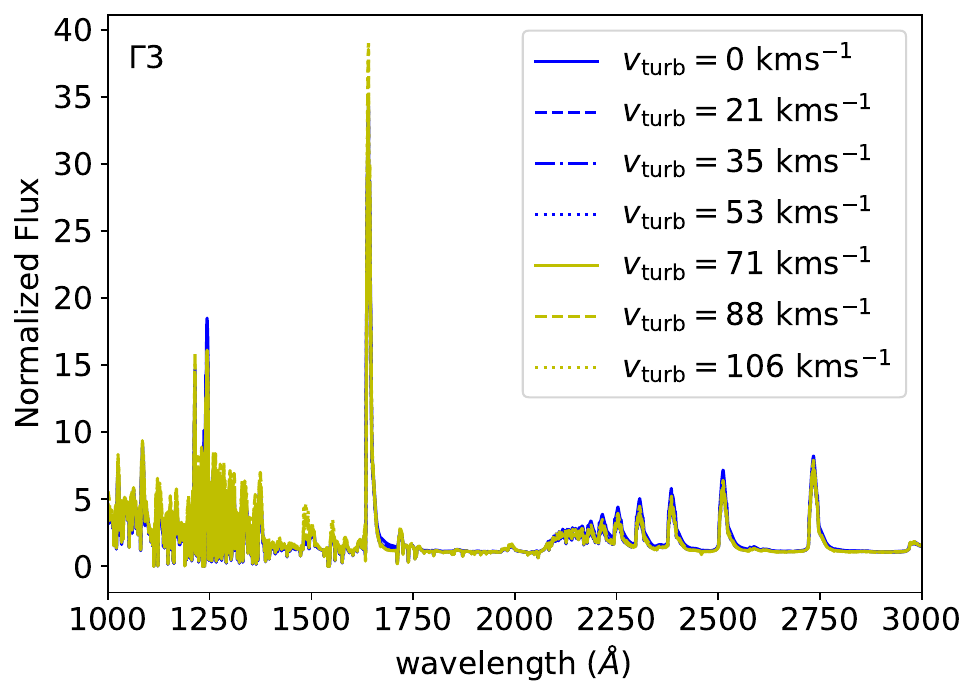}
      \caption{  Computed UV spectra for the PoWR$^{\textsc{hd}}$ $\Gamma$3 models including different  $\varv_{\mathrm{turb}}$ values.
      }
      \label{fig:iuevturb}
  \end{figure}

\begin{figure}
      \centering
      \includegraphics[width=1.\linewidth]{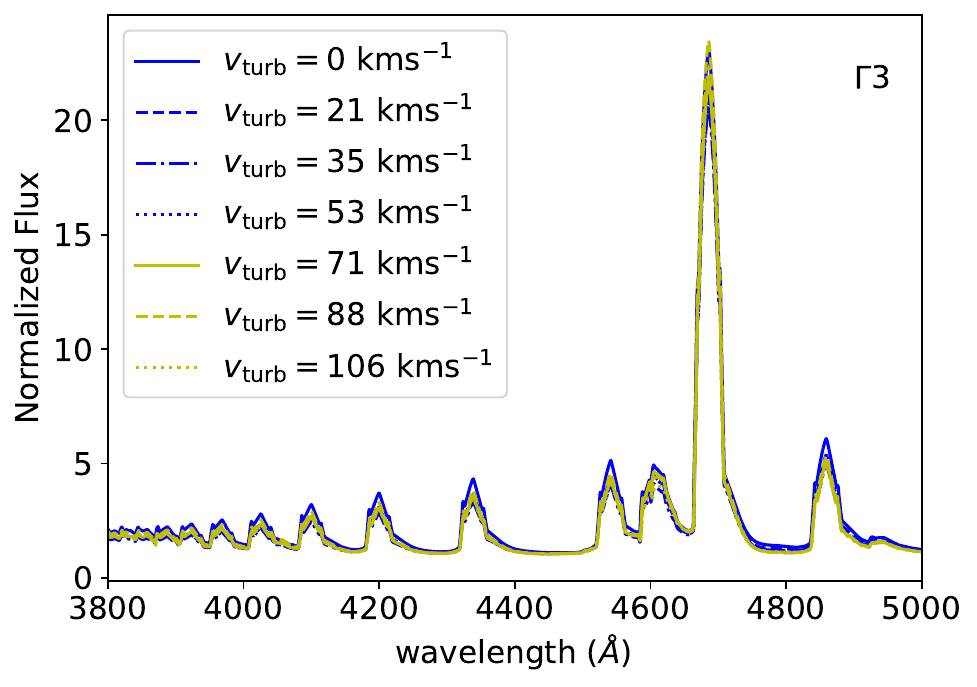}
      \caption{ Same as Fig.~\ref{fig:iuevturb} for the Optical spectral range.
      }
      \label{fig:optvturb}
  \end{figure}

Figures~\ref{fig:iuevturb} and \ref{fig:optvturb} show the UV and optical spectral regions for the PoWR$^{\textsc{hd}}$ with fixed stellar mass, luminosity and $\varv_{\mathrm{Dop}}=50\,\mathrm{km}\,\mathrm{s}^{-1}$ but varying turbulence. A zoomed-in version for the UV spectra is shown in Figure~\ref{fig:iuevturb2}. We see that unlike for the O stars in \citet{GonzalezTora+25}, the change in turbulence does not significantly affect the spectral lines as the impact concerns the broadening of lines in the quasi-hydrostatic regime while all the lines here are formed in the wind.

\begin{figure}
      \centering
      \includegraphics[width=1.\linewidth]{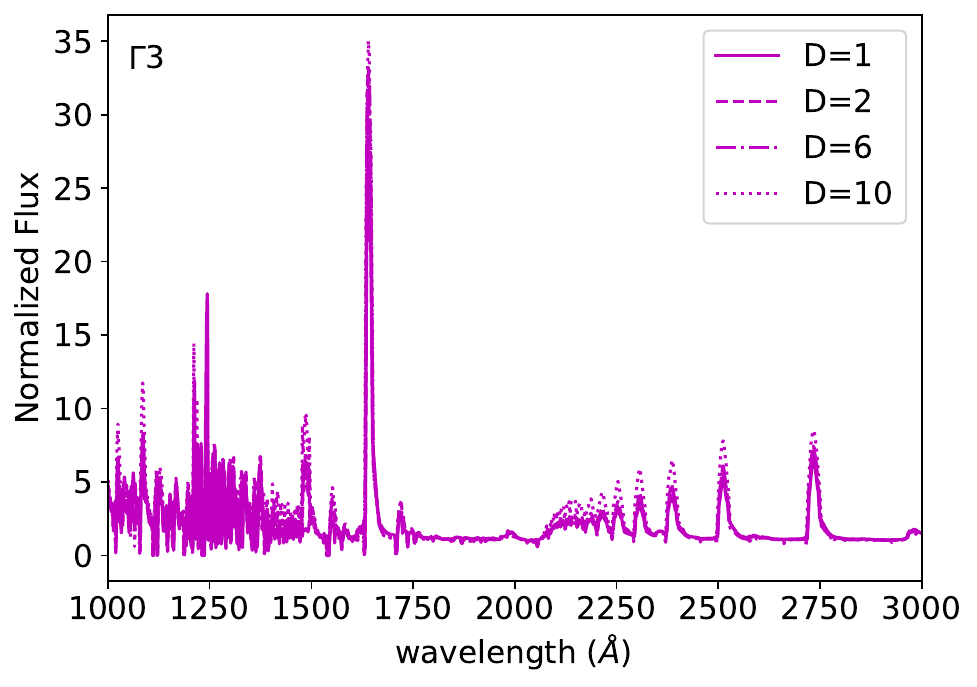}
      \caption{  Computed UV spectra for the PoWR$^{\textsc{hd}}$ $\Gamma$3 models including different density contrast values.
      }
      \label{fig:iuedenscon}
  \end{figure}

\begin{figure}
      \centering
      \includegraphics[width=1.\linewidth]{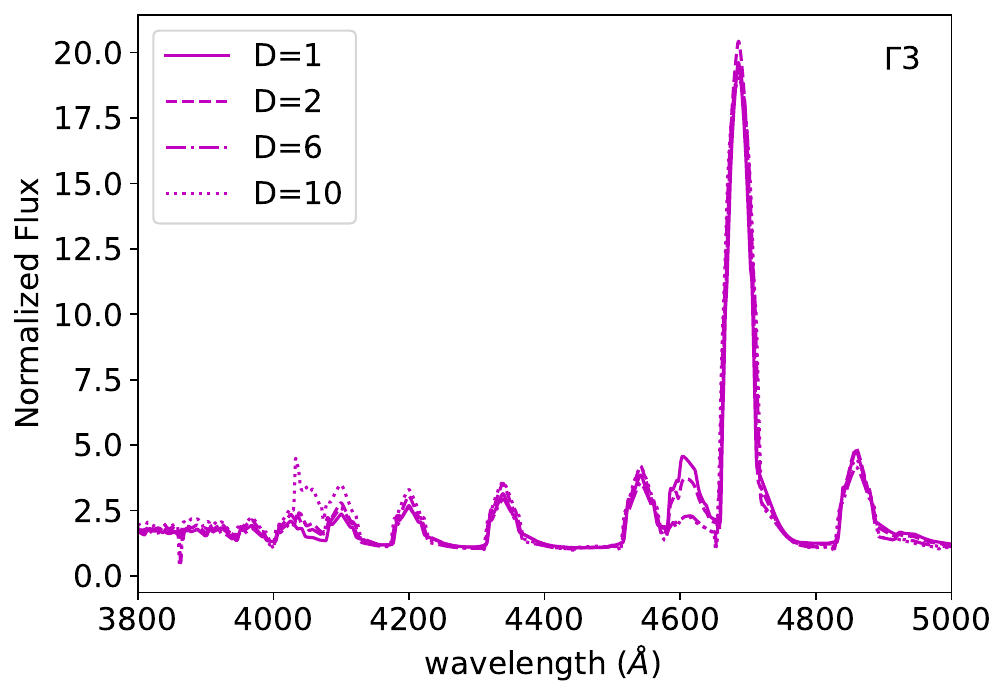}
      \caption{ Same as Fig.~\ref{fig:iuedenscon} for the Optical spectral range.
      }
      \label{fig:optdenscon}
  \end{figure}

Figures~\ref{fig:iuedenscon} and \ref{fig:optdenscon}  show the UV and Optical synthetic spectra from the models with different density contrasts, from a smooth model $D=1$ to a clumped model with $D=10$ following a clumping law from \citet{Hillier+2003} and a characteristic velocity of 100 km\,s$^{-1}$. A zoomed-in version for the UV spectra is shown in Figure~\ref{fig:iuedenscon2}. While the UV spectra remain relatively unchanged, the optical spectra present a difference in \ion{N}{v}$\,4604\,$\AA\ and \ion{N}{iii}$\,4640\,$\AA\, and to a lesser extent the \ion{N}{iv}$\,4057\,$\AA\ line because of the high terminal velocities in the outer wind. 

Despite the slight changes on the optical spectra in Figs.~\ref{fig:opt} and \ref{fig:optdenscon}, the optical regime does not fully reflect the differences in wind characteristics and can ultimately be prone to degeneracies in parameter determination.
In this case, as pointed out by \citet{Lefever+2023}, the UV yield much better diagnostics as the extent of wind differences is only fully reflected in well-resolved P-Cygni lines such as \ion{C}{iv}$\,1550\,$\AA. 
To illustrate the wind differences, Figures~\ref{fig:pcygpowr}, \ref{fig:pcygvturb} and  \ref{fig:pcygdenscon} zoom in on the UV \ion{C}{iv}$\,1550\,$\AA\, line, corresponding to the models shown in Figs.~\ref{fig:iue}, \ref{fig:iuevturb}, and \ref{fig:iuedenscon}, respectively. In Fig.~\ref{fig:pcygpowr} for all the $\Gamma$2, $\Gamma$3 and $\Gamma$4 profiles, the blue-most part of the absorption troughs shift significantly in accordance with wind parameter differences of the various models (see, e.g., $\varv_\infty$ in Table~\ref{tab:paramsbeta}). In Fig.~\ref{fig:pcygdenscon}, the different clumping parameters will also affect the P-Cygni line, as expected.  
In contrast, there is very little variation in the P-Cygni line in Fig.~\ref{fig:pcygvturb} which illustrates the changes in $\varv_{\mathrm{turb}}$. This little variation is expected with the comparatively small wind changes in Figs.~\ref{fig:vturbmdot} and \ref{fig:profilevturb}. 

%-----------------------------------------------------------------

\section{Conclusions}\label{sec:conc}

We have compared two different complementary hydrodynamical modelling approaches to reproduce the atmospheric layers and stellar outflows of three cWR stars: the 1D, spherically symmetric, stationary and non-LTE PoWR$^{\textsc{hd}}$ framework and the 3D, box-in-a-star, time-evolution, LTE models by \citet{Moens+2022wr}. Albeit much needed, it is currently computationally unfeasible for atmospheric models to solve the complex interplay between radiation field and atomic physics with 3D and time-dependent simulations. Therefore, using and comparing these two complementary modeling approaches is paramount to retrieve valuable knowledge on the atmospheres of WRs.  

Generally, we obtain a very good agreement between the two methods. The 1D models manage to reproduce the overall behavior of the averaged velocity, density, and temperature stratification of the 3D models. The density is best reproduced, while the velocities are a bit overestimated and temperatures in the outer wind are lower in 1D models, which is similar to our O-star comparison and likely a consequence of the different treatments \citep[see also][]{GonzalezTora+25}. For the same stellar masses and luminosities as in \citet{Moens+2022wr}, the 1D PoWR$^{\textsc{hd}}$ models overestimate $\log\dot{M}$ by up to $\lesssim0.2$ dex and $\varv(6\,R_\ast)$ by up to $\sim400\,\mathrm{km\,s}^{-1}$. Consequently, the resulting effective temperatures are lower by $0.2\,$dex ($\sim$$30$kK). These discrepancies can largely be mitigated by small adjustments of the mass ($<2\%$) or the luminosity ($< 0.1\,$dex) of the 1D models. Such differences are in agreement with the mass loss and luminosity dispersions obtained by the 3D simulations. 

We further explored the effect on varying additional parameters inherent to hydrodynamically consistent 1D PoWR$^{\textsc{hd}}$ models, namely clumping, Doppler velocity, and turbulent pressure, and compared with a $\beta$-law model. Clumping usually worsens the agreement with the 3D average velocity, which is expected given the low clumping amount in the \citet{Moens+2022wr} models. Decreasing the Doppler velocity in the PoWR models from $100\,\mathrm{km\,s}^{-1}$ to $50\,\mathrm{km\,s}^{-1}$ can help reconcile the discrepancies between both 1D and 3D approaches, but the necessary Doppler velocity seems to be regime-dependent with models further away from the Eddington limit requiring a higher Doppler velocity. The inclusion of a larger, constant turbulent pressure in the models slightly increases the derived mass-loss rates and also the terminal velocities.

Using the ability of PoWR to perform spectral synthesis, we compare the resulting model spectra of the different 1D PoWR$^{\textsc{hd}}$ representations and a $\beta$-law model adopting the main 3D model parameters. While the $\beta$-law branch overall matches the PoWR$^{\textsc{hd}}$ spectra for the optical part, notable differences occur for the \ion{N}{v}$\,4604\,$\AA\ and \ion{N}{iii}$\,4640\,$\AA\ lines due to the different ionization states between the $\Gamma$2 PoWR$^{\textsc{hd}}$ and the $\beta$-law approaches.

The $\Gamma$2, $\Gamma$3 and $\Gamma$4 models reproduced here launch a wind in the subsurface regions of a supersonic optically thick regime and are well reproduced by the 1D PoWR models, enabling to reasonably perform spectral analysis of cWR stars with dynamically-consistent models in future work. Yet, our models have problems with the lower luminosity $\Gamma$1 model in \citet{Moens+2022wr}, simulating an inflated and turbulent atmosphere with an optically thin line-driven wind on top, which corresponds more to a transition from a WR to a stripped hot (sub)dwarf. We could not obtain a converged model for the $\Gamma$1 model with the current version of PoWR$^\textsc{hd}$, obtaining unrealistically low $\dot{M}$ values. This is likely the result of the current inability to reproduce the turbulent environment with a subsequent wind launching. Further updates of the hydrodynamic treatment in 1D atmospheres, such as an improved, depth-dependent treatment of turbulent pressure could help bridge the regime between hot stripped stars and cWR stars.

\begin{acknowledgements}
GGT is supported by the German Deutsche Forschungsgemeinschaft (DFG) under Project-ID 496854903 (SA4064/2-1, PI Sander). 
GGT further acknowledges financial support by the Federal Ministry for Economic Affairs and Climate Action (BMWK) via the German Aerospace Center (Deutsches Zentrum f\"ur Luft- und Raumfahrt, DLR) grant 50 OR 2503 (PI: Sander).
AS, MBP, RRL, and JJ are supported by the German Deutsche Forschungsgemeinschaft (DFG) under Project-ID 445674056 (Emmy Noether Research Group SA4064/1-1, PI Sander). GGT and AS further acknowledge support from the Federal Ministry of Education and Research (BMBF) and the Baden-W{\"u}rttemberg Ministry of Science as part of the Excellence Strategy of the German Federal and State Governments. JOS, DD, LD, NM, CVdS acknowledge the support of the European Research Council (ERC) Horizon Europe grant under grant agreement number 101044048 (ERC-2021-COG, SUPERSTARS-3D) and from KU Leuven C1 grant BRAVE C16/23/009. OV, JOS, and LD acknowledge the support of the Belgian Research Foundation Flanders (FWO) Odysseus program under grant number G0H9218N and FWO grant G077822N.
\end{acknowledgements}

   \bibliographystyle{aa} % style aa.bst
   \bibliography{sample} % your references Yourfile.bib

 \begin{appendix}
\section{Gas, radiative and flux temperatures}\label{ap:trad}
\mbox{}
       \begin{figure}[h]
      \centering
      \includegraphics[width=.85\linewidth]{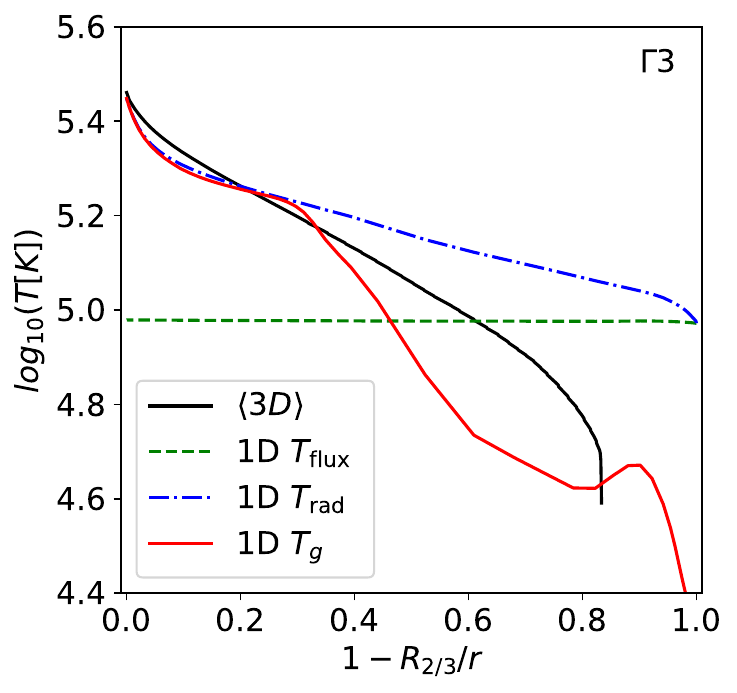}
      \caption{Flux temperature (dashed green line) and radiation temperature (dashed blue line) for the $\Gamma$3 model obtained from the total emergent flux, compared to the gas temperature stratifications for the averaged $\langle$3D$\rangle$ model. 
      }
      \label{fig:trad}
  \end{figure}
\section{$\log\dot{M}$ vs $\log\varv_{\mathrm{Dop}}$ scaling}\label{ap:slope}
\mbox{}
\begin{figure}[h]
      \centering
      \includegraphics[width=.85\linewidth]{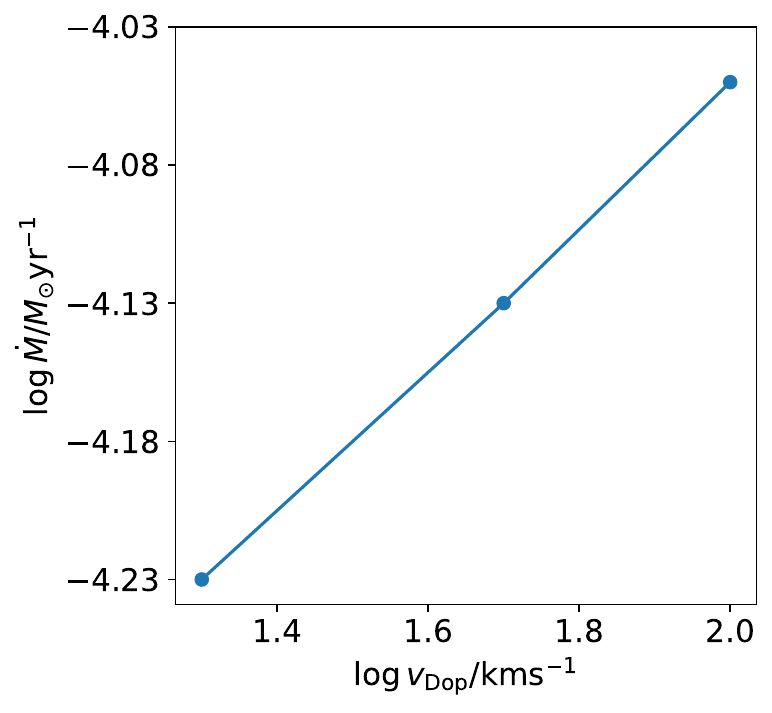}
      \caption{ Obtained $\log\dot{M}$ with respect to the assumed $\log\varv_{\mathrm{Dop}}$ for the $\Gamma$3 models. 
      }
      \label{fig:slope}
  \end{figure}
%\hfill \break
%\columnbreak
%\vfill\null
%\vfill\null
%\columnbreak
%\section{The UV \ion{C}{iv} $ - 1550\,$\AA\ P-Cygni profile}\label{ap:pcyg}
 %      \begin{figure}[h]
  %    \centering
   %   \includegraphics[width=.95\linewidth]{Figures/Moens22b-spectra-Pcyg-tmf2.pdf}
    %  \caption{ The UV \ion{C}{iv} $ - 1550\,$\AA\ P-Cygni profile for the PoWR models.  \textit{From up to down:} the $\beta$-law approach (blue solid line), the PoWR$^{\textsc{hd}}$ with identical input as \citet{Moens+2022wr} (red dashed dotted), the PoWR$^{\textsc{hd}}$ with flexible mass (red dotted line) and the PoWR$^{\textsc{hd}}$ with flexible luminosity (red dash-dotted line), for $\Gamma$2, $\Gamma$3 and $\Gamma$4 models, respectively.
     % }
      %\label{fig:pcygpowr}
  %\end{figure}

   %      \begin{figure}[h]
    %  \centering
     % \includegraphics[width=.95\linewidth]{Figures/Gamma3_vturb_spectra-Pcyg-vd50-tmf.pdf}
     % \caption{ The UV \ion{C}{iv} $ - 1550\,$\AA\ P-Cygni profile for the PoWR$^{\textsc{hd}}$ $\Gamma$3 models including a different $\varv_{\mathrm{turb}}$ values.
     % }
     % \label{fig:pcygvturb}
  %\end{figure}

   %      \begin{figure}[h]
    %  \centering
     % \includegraphics[width=.95\linewidth]{Figures/Gamma3-denscon-spectra-Pcyg-tmf2.pdf}
      %\caption{ The UV \ion{C}{iv} $ - 1550\,$\AA\ P-Cygni profile for the PoWR$^{\textsc{hd}}$ $\Gamma$3 models including a different density contrast values. 
      %}
      %\label{fig:pcygdenscon}
  %\end{figure}
%\columnbreak
\newpage \onecolumn
\section{Zoomed-in UV spectra}\label{ap:uvspectra}
\mbox{}
\begin{figure*}[h]
      \centering
      \includegraphics[width=1.\linewidth]{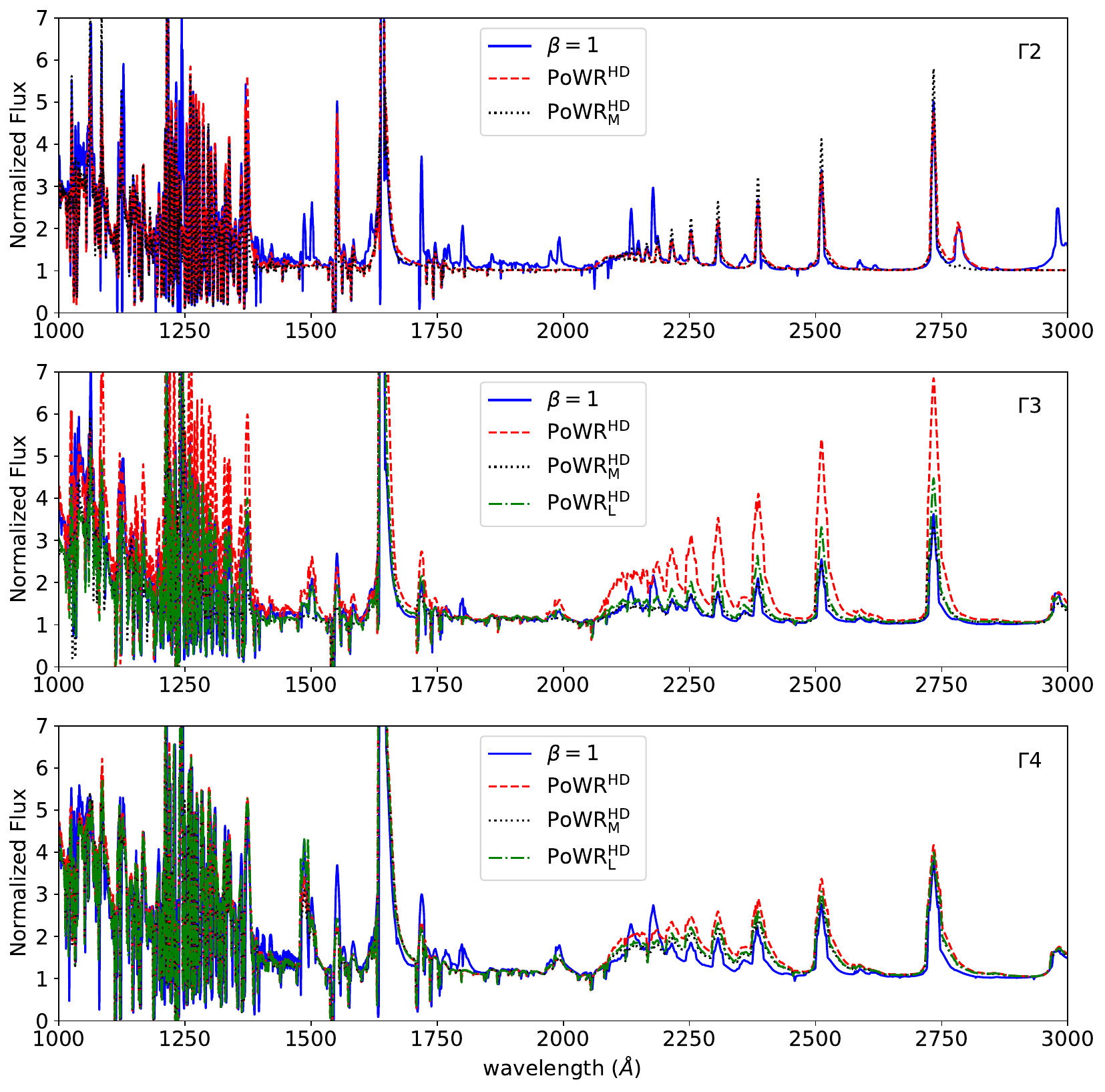}
      \caption{ Zoomed-in of the computed UV spectra for the PoWR models. \textit{From up to down:} the $\beta$-law approach (blue solid line), the PoWR$^{\textsc{hd}}$ with identical input as \citet{Moens+2022wr} (red dashed dotted), the PoWR$^{\textsc{hd}}$ with flexible mass (red dotted line) and the PoWR$^{\textsc{hd}}$ with flexible luminosity (red dash-dotted line), for $\Gamma$2, $\Gamma$3 and $\Gamma$4 models, respectively.
      }
      \label{fig:iue2}
  \end{figure*}

  \begin{figure*}[h]
      \centering
      \includegraphics[width=1.\linewidth]{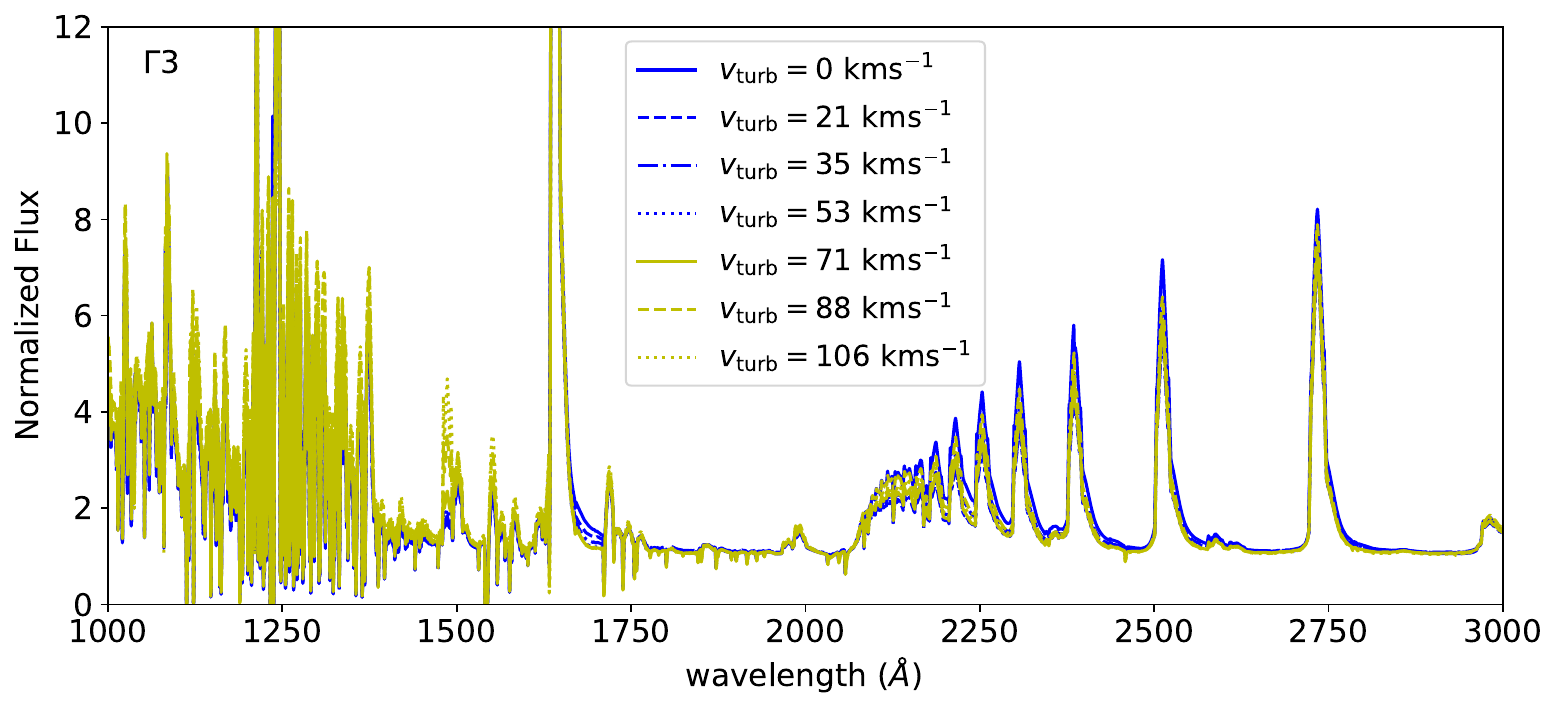}
      \caption{  Zoomed-in of the computed UV spectra for the PoWR$^{\textsc{hd}}$ $\Gamma$3 models including different  $\varv_{\mathrm{turb}}$ values.
      }
      \label{fig:iuevturb2}
  \end{figure*}

   \begin{figure*}[h]
      \centering
      \includegraphics[width=1.\linewidth]{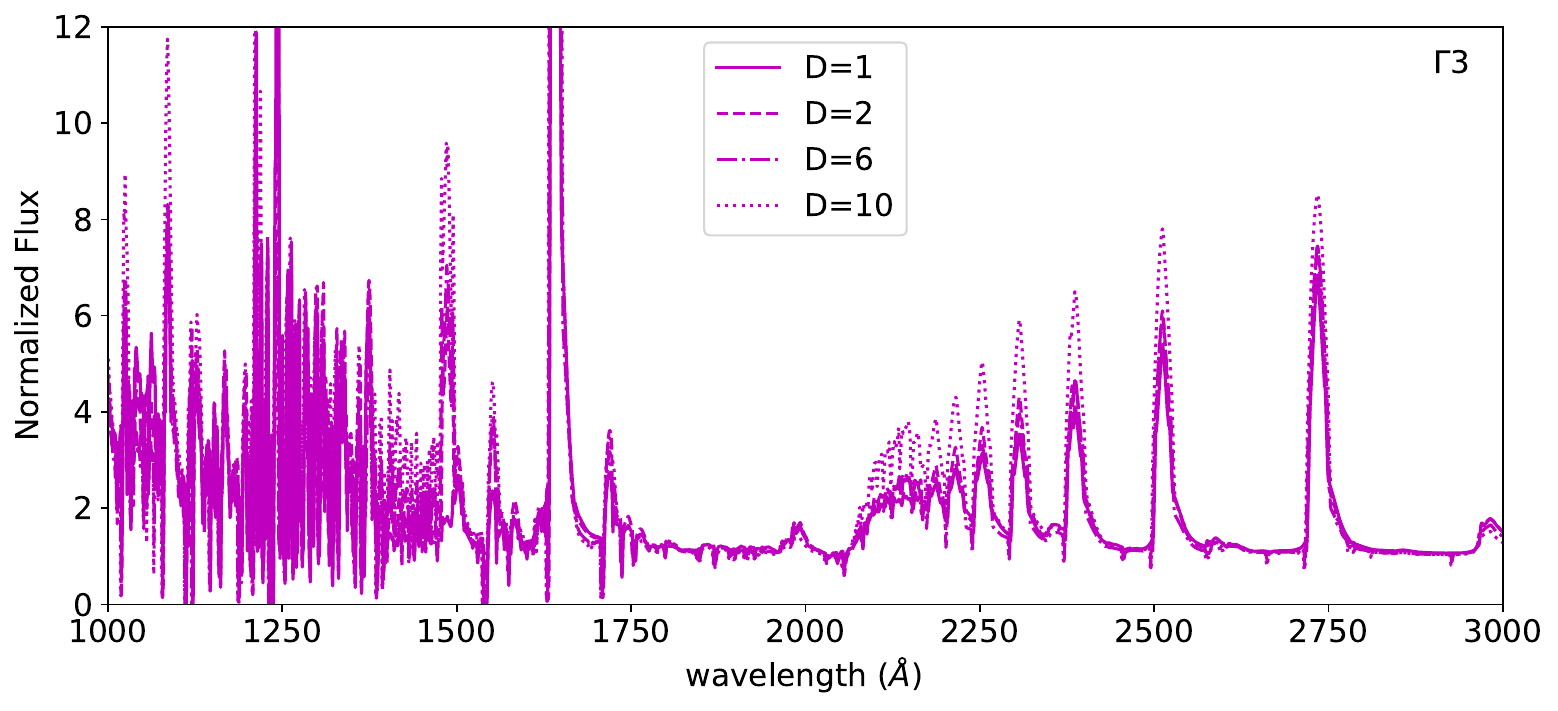}
      \caption{  Zoomed-in of the computed UV spectra for the PoWR$^{\textsc{hd}}$ $\Gamma$3 models including different density contrast values.
      }
      \label{fig:iuedenscon2}
  \end{figure*}
\newpage \twocolumn
\section{The UV \ion{C}{iv} $ - 1550\,$\AA\ P-Cygni profile}\label{ap:pcyg}
       \begin{figure}[h]
      \centering
      \includegraphics[width=.95\linewidth]{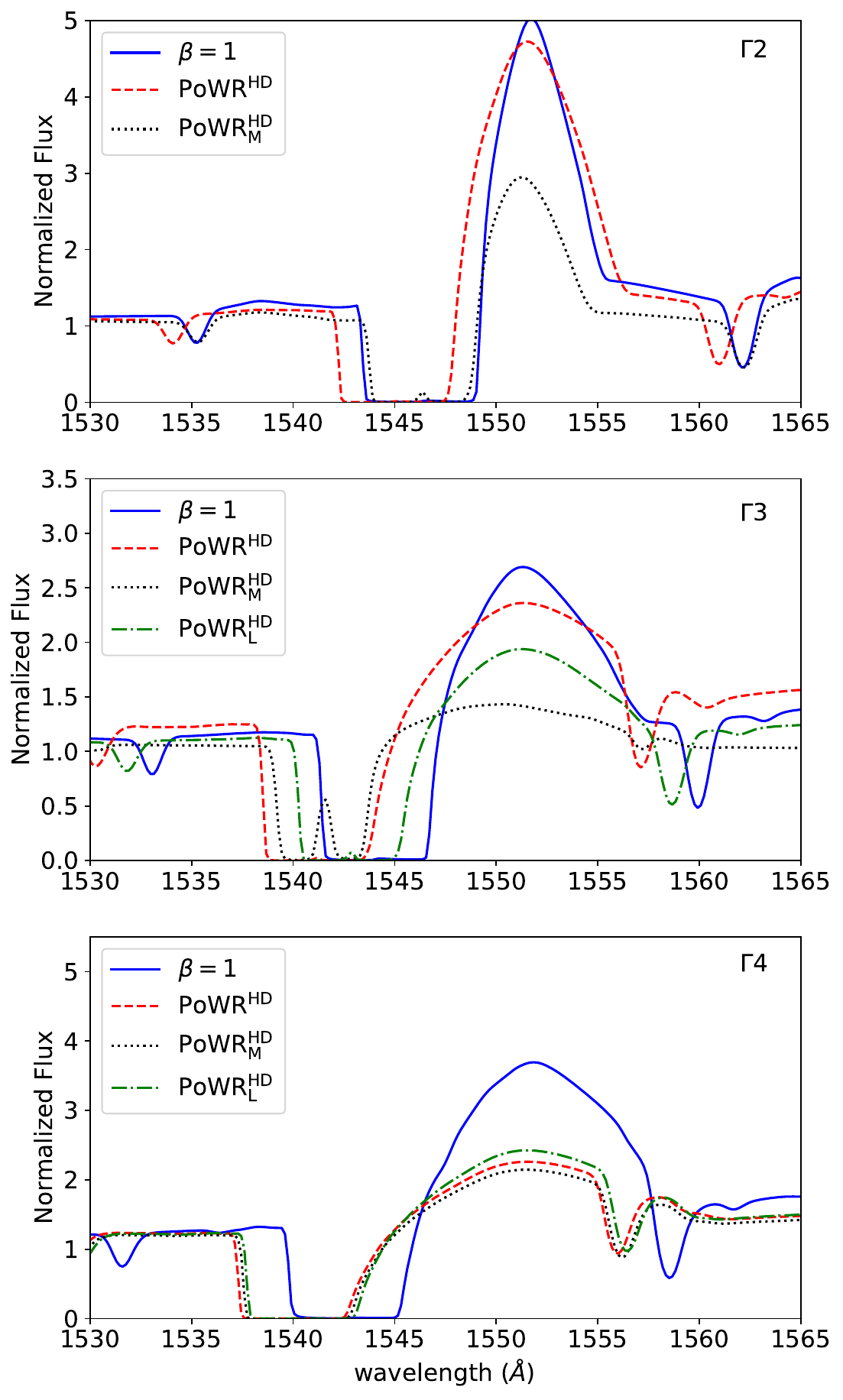}
      \caption{ The UV \ion{C}{iv} $ - 1550\,$\AA\ P-Cygni profile for the PoWR models.  \textit{From up to down:} the $\beta$-law approach (blue solid line), the PoWR$^{\textsc{hd}}$ with identical input as \citet{Moens+2022wr} (red dashed dotted), the PoWR$^{\textsc{hd}}$ with flexible mass (red dotted line) and the PoWR$^{\textsc{hd}}$ with flexible luminosity (red dash-dotted line), for $\Gamma$2, $\Gamma$3 and $\Gamma$4 models, respectively.
      }
      \label{fig:pcygpowr}
  \end{figure}

         \begin{figure}[h]
      \centering
      \includegraphics[width=.95\linewidth]{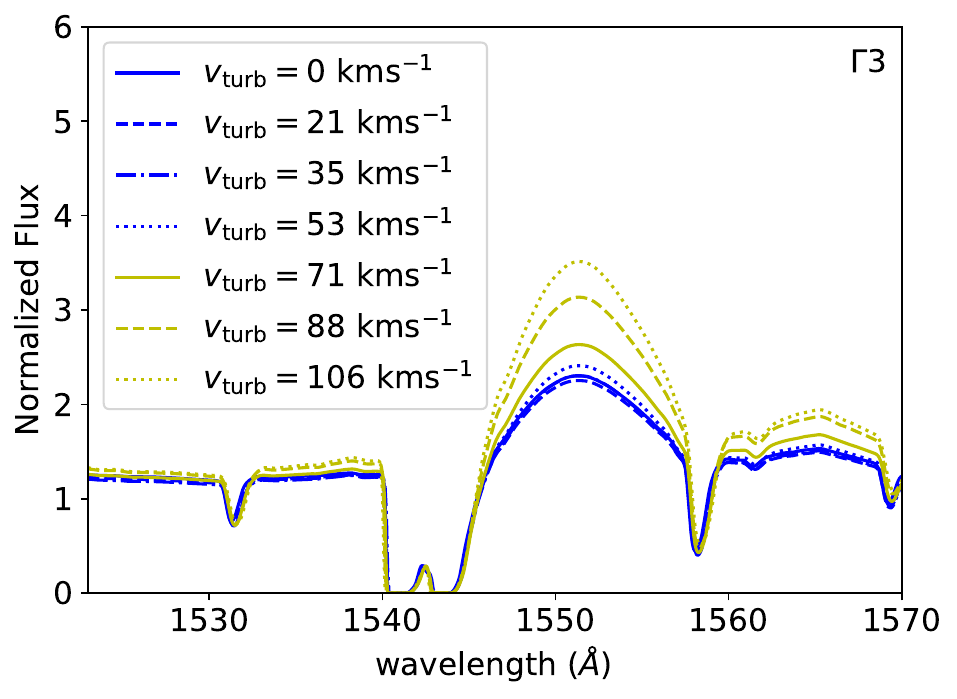}
      \caption{ The UV \ion{C}{iv} $ - 1550\,$\AA\ P-Cygni profile for the PoWR$^{\textsc{hd}}$ $\Gamma$3 models including a different $\varv_{\mathrm{turb}}$ values.
      }
      \label{fig:pcygvturb}
  \end{figure}

         \begin{figure}[h]
      \centering
      \includegraphics[width=.95\linewidth]{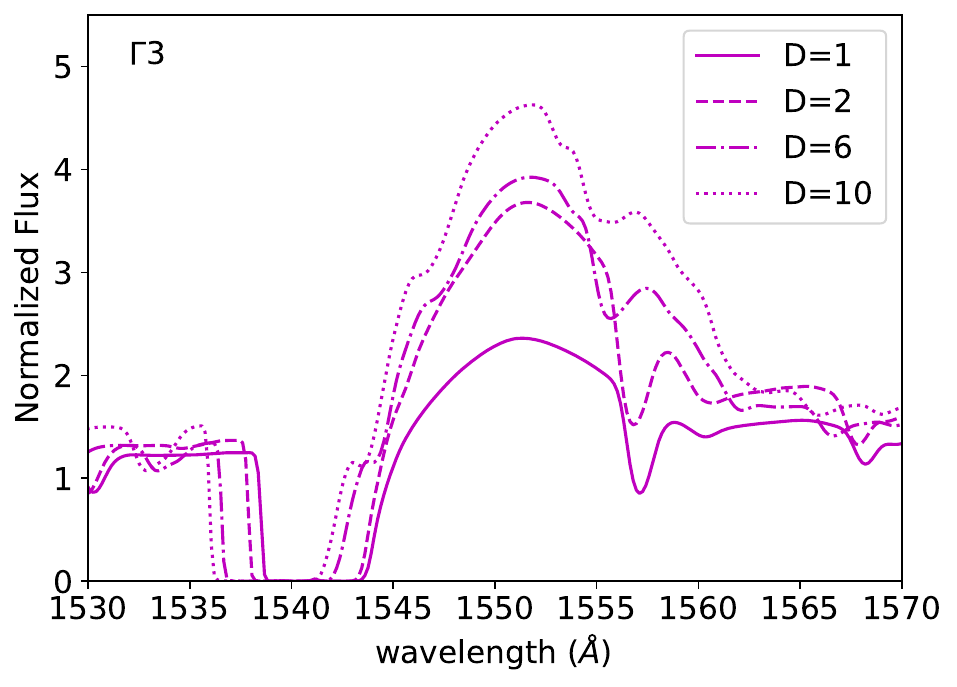}
      \caption{ The UV \ion{C}{iv} $ - 1550\,$\AA\ P-Cygni profile for the PoWR$^{\textsc{hd}}$ $\Gamma$3 models including a different density contrast values. 
      }
      \label{fig:pcygdenscon}
  \end{figure}

\end{appendix}
\end{document}